\documentclass[aps,prd,onecolumn,showpacs,eqsecnum,nofootinbib]{revtex4}
     
\usepackage{epsfig}
\usepackage{graphicx}
\usepackage{dcolumn}
\usepackage{amsmath}
\usepackage{enumerate}

\def\beq{\begin{equation}}
\def\eeq{\end{equation}}

\def\gtap{\mathrel{ \rlap{\raise 0.511ex \hbox{$>$}}{\lower 0.511ex
   \hbox{$\sim$}}}} 
\def\ltap{\mathrel{ \rlap{\raise 0.511ex
    \hbox{$<$}}{\lower 0.511ex \hbox{$\sim$}}}} 
\newcommand{\bea}{\begin{eqnarray}} \newcommand{\eea}{\end{eqnarray}}
\newcommand{\deltaunodue}{\mbox{$\Delta m_{21}^2 $}}
\newcommand{\deltaunotre}{\mbox{$\Delta m_{31}^2 $}}

\newcommand{\tetaot}{\mbox{$\theta_{13}$}}

\newcommand{\pbarP}{P \hspace{-3mm}
\raisebox{1.9ex}{{\tiny(}}\raisebox{1.4ex}{--}\raisebox{1.9ex}{\tiny)}
   }
\newcommand{\sensP}{\mbox{${\cal S}(\delta)$}}
\newcommand{\sensPbar}{\mbox{${\cal \bar{S}}(\delta)$}}

\begin{document}

\vskip-6pt \hfill {IPPP/06/85} \\
\vskip-6pt \hfill {DCPT/06/170} \\
\vskip-6pt \hfill {FERMILAB-PUB-07-021-E} \\
\vskip-6pt \hfill {Roma-TH-1443} \\
\title{
\vskip-12pt~\\
A low energy neutrino factory for large $\theta_{13}$}

\author{\mbox{Steve Geer$^{1}$}
\mbox{Olga Mena$^{2,3}$} and
\mbox{Silvia Pascoli$^{4}$}}

\affiliation{
\mbox{$^2$ Theoretical Physics Department, Fermi National Accelerator
Laboratory, Batavia, IL 60510-0500, USA}
\mbox{$^3$ INFN - Sez. di Roma,                    
Dipartimento di Fisica,                 
Universit\'a di Roma "La Sapienza",      
P.le A. Moro, 5,                       
I-00185 Roma,
Italy}
\mbox{$^4$ IPPP, Department of Physics, Durham University, Durham
  DH1 3LE, United Kingdom}
\\
{\tt omena@fnal.gov},
{\tt silvia.pascoli@durham.ac.uk}
}

\begin{abstract}

If the value of $\theta_{13}$ is within the reach of the upcoming 
generation of long-baseline experiments,
T2K and NO$\nu$A, we show that  
a low-energy neutrino factory, with peak energy in the few GeV range,
 would provide a sensitive tool to explore CP-violation and the neutrino mass hierarchy.
 We consider baselines with typical length 1000--1500~km.
The unique performance of the low energy neutrino factory
 is due to the rich neutrino oscillation pattern at energies between $1$ and $4$ GeV at baselines $\mathcal{O}(1000)$ km. 
 We perform both a semi-analytical study of the sensitivities 
 and a numerical analysis to explore how well this setup can measure 
 $\theta_{13}$, CP-violation, and determine the type of mass hierarchy and the $\theta_{23}$  quadrant.
 A  \emph{low energy neutrino factory} provides a powerful tool to resolve ambiguities and make precise parameter 
determinations, for both large and fairly small values of the mixing 
parameter $\theta_{13}$.

\end{abstract}


\maketitle

\section{Introduction}
During the last several years our understanding of the physics of neutrinos has made remarkable progress. The experiments with solar~\cite{sol,SKsolar,SNO1,SNO2,SNO3,SNOsalt},  atmospheric~\cite{SKatm}, reactor~\cite{KamLAND}, and also
long-baseline accelerator~\cite{K2K} neutrinos have provided compelling evidence for the existence of neutrino oscillations, implying non zero neutrino
masses. The present data\footnote{We
  restrict ourselves to a three-family neutrino analysis. The
  unconfirmed LSND signal~\cite{LSND} cannot be explained within this
  context and might require additional light sterile neutrinos or more
  exotic explanations. The ongoing
  MiniBooNE experiment~\cite{miniboone} is going to test the
  oscillation explanation of the LSND result.} require two large
($\theta_{12}$ and $\theta_{23}$) and one small ($\theta_{13}$) angles
in the lepton mixing matrix~\cite{BPont57}, and at least two mass squared differences,
$\Delta m_{ji}^{2} \equiv m_j^2 -m_i^2$ (where $m_{j}$'s are the neutrino
masses), one driving the atmospheric ($\deltaunotre$) and the other one the solar ($\deltaunodue$) neutrino oscillations. The mixing
angles $\theta_{12}$ and $\theta_{23}$ control the solar and the
dominant atmospheric neutrino oscillations, while $\theta_{13}$ is the
angle constrained by the data from the CHOOZ and Palo Verde reactor
experiments~\cite{CHOOZ,PaloV}.

The Super-Kamiokande~\cite{SKatm} and K2K~\cite{K2K} data are well
described in terms of dominant $\nu_{\mu} \rightarrow \nu_{\tau}$
($\bar{\nu}_{\mu} \rightarrow \bar{\nu}_{\tau}$) vacuum
oscillations. The MINOS Collaboration~\cite{MINOS} has recently reported the first neutrino oscillation results from $1.27 \times 10^{20}$~\cite{MINOSRECENT}. The value of the oscillation parameters from MINOS are consistent with the ones from K2K, as well as from SK data. A recent global fit~\cite{thomas} (see also Ref.~\cite{newfit}) provides the following $3 \sigma$ allowed ranges for the atmospheric mixing parameters
\beq 
\label{eq:range}|\deltaunotre| =(1.9 - 3.2)\times10^{-3}\ {\rm eV^2},~~~~
0.34<\sin^2\theta_{23}<0.68~.
\eeq
The sign of $\deltaunotre$, sign$(\deltaunotre)$, 
cannot be determined with the existing data. The two possibilities,
$\deltaunotre > 0$ or $\deltaunotre < 0$, correspond to two different
types of neutrino mass ordering: normal hierarchy and inverted hierarchy. In addition, information on the octant in which $\theta_{23}$ lies, if $\sin^22\theta_{23} \neq 1$, is beyond the reach of present experiments. 

The 2-neutrino oscillation analysis of the solar neutrino data,
including the results from the complete salt phase of the Sudbury
Neutrino Observatory (SNO) experiment~\cite{SNOsalt}, in combination
with the KamLAND spectrum data~\cite{KL766}, shows that the solar neutrino oscillation parameters lie in the low-LMA (Large Mixing Angle) region, with best fit values~\cite{thomas} $\deltaunodue =7.9 \times 10^{-5}~{\rm eV^2}$ and $\sin^2 \theta_{12} =0.30$.

A combined 3-neutrino oscillation analysis of the solar, atmospheric,
reactor and long-baseline neutrino data~\cite{thomas} constrains the third mixing angle to be $\sin^2\theta_{13} < 0.041$ at the $3\sigma$ C.L. However, the bound on $\sin^2 \theta_{13}$ is dependent on the precise value of $|\Delta m^2_{31}|$,
being stronger for larger values of $|\Delta m^2_{31}|$.
The future goals for the study of neutrino properties in neutrino oscillation experiments
 is to
precisely determine the already measured oscillation parameters
and to obtain information on the unknown ones: namely $\theta_{13}$,
the CP--violating phase $\delta$ and the type of neutrino mass
hierarchy (or equivalently sign$(\deltaunotre)$).                                                          
 
In the next sections we will explore in detail the
 possible measurement of the two unknown parameters $\theta_{13}$ and $\delta$ with a future  {\it neutrino factory}~\cite{geer} facility as this appears to be among the most promising ways to unveil neutrino mixing and leptonic CP violation~\cite{nf1,nf2,nf3,nf4,nf5,nf6,nf6b,nf7,nf7b,nf7c,silver,study1-physics,nf8,yo,nf9}\footnote{For the prospects of future measurements of these two oscillation parameters at $\beta$ beam experiments~\cite{zucc, mauro}, see Refs.~\cite{betabeam1,betabeam2,betabeam3,betabeam4,betabeam5,tabarellis1}.}.

A neutrino factory consists of a high intensity muon 
source\footnote{A neutrino factory muon source is an attractive 
stepping-stone towards a high energy muon collider.}, an acceleration 
system, and a muon storage ring with long straight sections. Muons decaying 
along the straight sections create high intensity neutrino beams which have 
a precisely-known flux, divergence, energy spectrum, and neutrino flavor 
content. The flavor composition of the beam depends on whether positive or negative muons 
are stored in the ring. Suppose, for example, that positive charged muons 
have been stored. Muons decaying in the straight sections will produce beams 
containing 50\% muon-antineutrinos and 50\% electron-neutrinos: $\mu^{+}\to 
e^{+} + \nu_e + \bar{\nu}_{\mu}$. Charged current interactions of the 
$\bar{\nu}_{\mu}$ in a distant detector will produce $\mu^+$ (``right-sign'' 
muons, i.e. muons of the same charge as those stored in the neutrino 
factory). In contrast, if the $\nu_e$'s oscillate to $\nu_\mu$'s and then 
interact in the far detector they will produce $\mu^-$ (``wrong-sign 
muons''~\cite{geer,nf1}). Thus, wrong-sign muons are an unambiguous proof of 
electron neutrino oscillation, in the $\nu_e\to\nu_\mu$ channel. This has 
been called the ``golden channel''~\cite{nf6}, and is central to the present 
study. A magnetized detector with excellent charge identification is 
necessary to exploit the golden channel.

In the following we will consider a \emph{low energy neutrino factory} where 
the stored muons have an energy of $4.12$ GeV.  This is motivated by recent 
progress in developing a viable concept for a neutrino factory detector with 
a threshold for reliably measuring the sign of muons with momenta down to a 
few hundred MeV/c~\cite{nufact06-det}. We explore the impact of analyzing 
the ``wrong-sign'' and ``right-sign'' muon rates for several energy bins, 
and consider two reference baselines: $1280$ Km, the distance from Fermilab 
to Homestake, and $1480$ Km, the distance from Fermilab to Henderson mine. 
Our results can be easily generalized to other baselines in the 1200--1500~km range.
We find that a simultaneous fit to the energy-dependent rates provides a 
powerful tool to resolve ambiguities and make precise parameter 
determinations, for both large and fairly small values of the mixing 
parameter $\theta_{13}$.

\section{Formalism}
\label{formalism}

In the present study we focus on the capabilities of 
a low-energy neutrino factory, if the value of $\theta_{13}$ 
is within reach of the upcoming 
generation of long-baseline experiments,
T2K and NO$\nu$A.

For neutrino energies $E \gtap$ 1 GeV, $\theta_{13}$ within
the present bounds~\cite{thomas,newfit}, and baselines $L \ltap
{\cal O} (1000)$~km~\cite{BMWdeg}, the oscillation
probability $\pbarP(L)$ can be expanded in the small parameters
$\theta_{13}$, $\Delta_{12}/\Delta_{13}$, $\Delta_{12}/A$ and
$\Delta_{12} L$ , where $\Delta_{12} \equiv \deltaunodue/(2 E)$ and
$\Delta_{13} \equiv \deltaunotre/(2E)$~\cite{nf6} (see also
Ref.~\cite{Akhmedov:2004ny}):
\beq
\begin{array}{ll}
\pbarP(L) \simeq &
\sin^2 \theta_{23} \, \sin^2 {2 \theta_{13} } \left(
\frac{\Delta_{13}}{A \mp \Delta_{13}} \right)^2
\sin^2 \left( \frac{(A \mp \Delta_{13}) L}{2} \right) \\
& + \cos \theta_{13} \sin {2 \theta_{13} } \sin {2 \theta_{23}} 
\sin {2 \theta_{12}} \ \frac{\Delta_{12}}{A} \frac{\Delta_{13}}{A \mp
  \Delta_{13}} \
\sin \left(\frac{A L}{2} \right) \sin \left( \frac{(A \mp
  \Delta_{13}) L}{2} \right)
 \cos \left(\frac{\Delta_{13} L}{2} \mp \delta \right) \\
 & + \cos^2 \theta_{23} \sin^2 {2 \theta_{12}} \left(
 \frac{\Delta_{12}}{A} \right)^2 \sin^2 \left( \frac{A L}{2} \right)~,
\end{array}
\label{eq:probappr}
\eeq
where the first, second and third terms have been dubbed \emph{atmospheric, interference and solar terms}, respectively.  
In the following analytical study,
we use the constant density approximation for the index of
refraction in matter $A \equiv \sqrt{2} G_{F} \bar{n}_e(L)$. Here,
$\bar{n}_e(L)= 1/L \int_{0}^{L} n_e(L') dL'$ is the average electron
number density, with $n_e(L)$ the electron number density along the
baseline.  

In order to study the sensitivity to CP-violation,
we introduce the weighted probability difference between the case of $\delta\neq 0$
and the one with no CP-violation $(\delta=0)$\footnote{ A similar analysis with similar results could be carried out for the case $\delta =\pi$. We present here the analytical study assuming that $\delta$ is within the interval $[-\pi/2,\pi/2]$. Our numerical simulations will consider the full $\delta$ range $[-\pi,\pi]$.}:
\begin{eqnarray}
\sensP    \equiv \frac{\Big( P(L, \delta) - P(L, 0) \Big)^2}{ P(L, \delta)} ~,\\
 \sensPbar    \equiv
 \frac{\Big( \bar{P}(L, \delta) - \bar{P}(L, 0) \Big)^2}{ \bar{P}(L, \delta)} ~.
 \label{eq:sensit}
 \end{eqnarray}
The quantity \sensP~(\sensPbar) 
is useful to get an estimate of the energy range, 
for a fixed baseline, for which the sensitivity is maximal. 
 Using Eq.~(\ref{eq:probappr}), we find that:
\begin{equation}
\sensP (\sensPbar) =\displaystyle \frac{ 4 \cos^2 \theta_{13} \sin^2 {2 \theta_{12}} \cos^2 \theta_{23} \Big(\displaystyle\frac{\Delta_{12}}{\Delta_{13} }\Big)^2 \Big( \displaystyle\frac{ \Delta_{13} L}{2} \Big)^2 
\Big( \cos ( \delta -  \displaystyle\frac{ \Delta_{13} L}{2} ) - \cos  \displaystyle\frac{ \Delta_{13} L}{2} \Big)^2}
{1 + 2 \epsilon \cos   ( \delta -  \displaystyle\frac{ \Delta_{13} L}{2} )  + \epsilon^2 }
\label{eq:sensP}
\end{equation}
where we have approximated $\cos \theta_{13} \simeq 1$. 
The quantity
$\epsilon$ is defined as:
\begin{equation}
\epsilon \equiv \frac{\cos \theta_{23}}{\sin \theta_{23}} \frac{\sin 2 \theta_{12}}{\sin 2\theta_{13}} \frac{\Delta_{12}}{\Delta_{13}} \frac{ \Delta_{13} L}{2} \frac{A\mp \Delta_{13}}{\Delta_{13}} \frac{1}{\sin ((A\mp \Delta_{13}) L/2)},
\end{equation}
where $\mp$ refers to neutrinos (antineutrinos) respectively. At leading order 
we can neglect $A/\Delta_{13}$ terms and $\epsilon$ is the same for neutrinos and antineutrinos.

The sensitivity to CP-violation is 0 when the interference term
in the oscillation probability,
e.g. the second term in the r.h.s. of Eq.~(\ref{eq:probappr}), 
cancels. For the values of $\theta_{13}$ of interest,
this happens at the
oscillation minima:
\begin{equation}
(\Delta_{13} \mp A) L/2= n  \pi \quad {n = 1,2,3...} ~.
\label{min1}
\end{equation}
Neglecting the small correction 
due to matter effects, we find that the minima in the sensitivity to CP-violation
corresponds to an energy of $E_m = 1.4 \ (1.2)$~GeV for L=1480 (1280)~km, with $n=1$. 
Here and in the following, we used 
$\Delta m^2_{31} = 2.4 \times 10^{-3}$~eV$^2$.
Matter effects modify this result by 10--20~\%.
At smaller energies additional minima and maxima with a fast oscillatory behaviour
are present but we will not consider 
such energy range due to detector resolutions, efficiencies and thresholds.
As at high energies \sensP~ and \sensPbar~ drop as $E^{-2}$,
we expect a maximum in sensitivity at an energy of few GeV. 
In the case of neutrinos and normal hierarchy (and, for negligible
$A/\Delta_{13}$, antineutrinos and inverted hierarchy),
an additional minimum in CP-violation sensitivity
is found for $\cos(\delta - \Delta_{13} L/2) = \cos ( \Delta_{13} L/2)$.
This equation has a solution for:
\begin{eqnarray}
\frac{\Delta_{13} L}{2} = \frac{\delta}{2} \ \ \ \  \ \ \ \ {\mathrm{for}}   \ \ \ \  \delta \geq 0  , \\
\frac{\Delta_{13} L}{2} = \pi - \frac{\delta}{2} \ \ \ \  \ \ \ \ {\mathrm{for}}   \ \ \ \  \delta <0 .
\label{min2}
\end{eqnarray}
For $\delta = \pi/2 \ (-\pi/2)$, the energy of the minimum in sensitivity is
at $E_m = 5.7 \ (1.9)$~GeV for $L=1480$~km and at
$E_m= 5.0  \ (1.7)$~GeV at $L=1280$~km.
Let us notice that, if $\delta < 0$, the two minima in Eqs.~(\ref{min1}) and (\ref{min2}) 
are very close and the maximum in between 
cannot be fully exploited due to limited energy resolution. 
The maximum in sensitivity at higher energy is located close
to the minimum in Eq.~(\ref{min1}), that is in the energy range
$E_M \sim (2$--$3)$~GeV, depending on the value of $\delta$.
For non-negative values of $\delta$, 
it is possible to make full use of the maximum
in sensitivity between the two minima in Eqs.~(\ref{min1}) and (\ref{min2}).
Again, the maximal sensitivity 
will be achieved at $E_M$ of few GeV.
In both cases, we can obtain a more precise estimate
for $E_M$ by neglecting $\epsilon$ and 
solving the equation
$ \Big( \cos ( \delta -  \frac{ \Delta_{13} L}{2} ) - \cos  \frac{ \Delta_{13} L}{2} \Big) + \frac{ \Delta_{13} L}{2} \Big (\sin ( \delta -  \frac{ \Delta_{13} L}{2} ) + \sin  \frac{ \Delta_{13} L}{2} \Big)=0$.
Notice that this equation holds only far away from the minima.
Typically, the maximum is reached in the energy range $(1.6$--5.2)~GeV [$(1.4$--4.5)~GeV] 
for $L=1480$~km [$L=1280$~km].
In particular, for $\delta = \pi/2 \ (- \pi/2)$ we have $E_M\simeq \, 1.7 \ (3.2)$~GeV [$1.5 \ (2.6)$~GeV]
with  $L=1480$~km [$L=1280$~km].
Matter effects included in $\epsilon$ in the denominator modify these results by a 10--20~\%.

For neutrinos and inverted hierarchy, (and antineutrinos and normal hierarchy, neglecting terms of order $A/\Delta_{13}$),
we find a similar behaviour.
The minimum in addition to the one
in Eq.~(\ref{min1}) is reached when $\cos(\delta + \Delta_{13} L/2) = \cos ( \Delta_{13} L/2)$
whose solutions are:
\begin{eqnarray}
\frac{\Delta_{13} L}{2} = \pi - \frac{\delta}{2} \ \ \ \  \ \ \ \ {\mathrm{for}}   \ \ \ \ \delta \geq 0, \\
\frac{\Delta_{13} L}{2} = \frac{\delta}{2}  \ \ \ \  \ \ \ \  {\mathrm{for}}  \ \ \ \  \delta <0 .
\end{eqnarray}
We can also estimate the position of the maximum by solving the equation
$ \Big( \cos ( \delta +\frac{ \Delta_{13} L}{2} ) - \cos  \frac{ \Delta_{13} L}{2} \Big) - \frac{ \Delta_{13} L}{2} \Big (\sin ( \delta +  \frac{ \Delta_{13} L}{2} ) - \sin  \frac{ \Delta_{13} L}{2} \Big)=0$.
The maximum is reached at approximately the same energy as for neutrinos but with opposite $\delta$. 
As far as $\epsilon$ is negligible, the sensitivity does not depend on $\theta_{13}$.
Our approximated analytical results on the energy maxima hold for 
$\sin^2 \theta_{13} \gg 10^{-3} \, (\Delta_{13} L /2)/ \sin ( \Delta_{13} L/2) $.
In conclusion, our analytical study
suggests that maximal sensitivity
to CP-violation is reached in the few GeV range.
Notice that, given one type of hierarchy, 
neutrinos and antineutrinos
have similar behaviour but for opposite 
values of $\delta$.
The combination of the two channels allows 
to reach optimal sensitivity 
independently of the true value
of the CP-violating phase.

Similar considerations hold also 
for the sensitivity to the type of hierarchy. 
We can study a similar quantity:
$(P_+ - P_-)^2 / P_+$ for neutrinos and antineutrinos. 
We find that the dominant term is proportional to $A L$ 
while CP-violating terms constitute a correction 
at most of 20\%--30\% for the highest allowed values of $\sin^2 2 \theta_{13}$. 
A minimum in the sensitivity is found in correspondence to the minima 
of the oscillation probability as in the case of CP-violation studied above.
The sensitivity to the type of hierarchy depends on the value of the 
$\delta$ phase once the CP-violating corrections are taken into account.
For $0 \leq \delta < \pi/2$, the energy for which a maximum in sensitivity is obtained, $E_M$, will be an decreasing function of the $\theta_{13}$ mixing angle.
For example, for $\delta=0$, the maximum in sensitivity will be reached
 in the 3.7--2.3~GeV range for $\sin^2 \theta_{13} = 0.01$--$0.1$.
 Conversely, for $\pi/2 < \delta \leq \pi$, $E_M$ will increase
 with $\theta_{13}$ but typically remain in the 1.3--1.6~GeV range. 
 We can conclude that maximal sensitivity is reached for
 energies around 1.3--4~GeV.
For antineutrinos, a similar behaviour can be found, with the exchange of
$\delta$ in $\pi - \delta$.

In order to study the sensitivity to CP-violation and type of hierarchy,
by exploiting the number of events in a simulated experiment,
the energy dependence of the cross sections and fluxes should be included.
We can expect the value of the energy for which the maxima of sensitivity are reached to be shifted slightly at higher values.

From the above considerations, we can conclude that the use of a detector 
with a low threshold and good energy resolution and efficiency at 
$E \gtap 1$~GeV is crucial for exploiting the potentiality of
a neutrino factory with baselines in the 1000-1500~Km. In addition,
a high energy neutrino beam is not necessary and 
it is sufficient to use lower energies with
respect to the commonly studied options for a neutrino factory
with muon energies of 20 or 50~GeV.

\section{The Low Energy Neutrino Factory concept}

A Neutrino Factory consists of (i) an intense low energy muon source, (ii) a 
muon beam manipulation and cooling system to maximize the number of muons 
within a useful acceptance, (iii) a ''pre-accelerator" to accelerate the 
muons from low kinetic energies (typically 100-200 MeV) to about 1 GeV, (iv) 
a system of one or more accelerators to further accelerate the muons to the 
desired final energy, and (v) a muon storage ring with long straight 
sections. Design studies~\cite{study1,study2,study2a} have shown that, for a 
20 GeV Neutrino Factory, the 1-20 GeV acceleration systems are expected to 
account for about $26\%$ of the estimated cost. Hence, if the physics goals 
can be met using muons with energies much lower than 20 GeV, there is a 
significant cost advantage. In the following, we first discuss the 
performance of the far detector (which places a lower limit on the desired 
Neutrino Factory muon energy), and then discuss the low energy Neutrino 
Factory and its performance. The primary neutrino oscillation channel at a 
Neutrino Factory requires identification of wrong-sign muons, and hence a 
detector with excellent muon charge identification. To obtain the required 
event rates, the far detector fiducial mass needs to be at least O(10 Kt), 
and therefore we require a very large magnetized detector. Early 
studies~\cite{study1-physics} based on a MINOS-like segmented magnetized 
detector suggested that, to reduce the charge mis-identification rate to the 
$10^{-4}$ level while retaining a reasonable muon reconstruction efficiency, 
the detected muon needs to have a minimum momentum of 5 GeV. The analysis 
obtained a 50\% reconstruction efficiency for CC neutrino interactions 
exceeding $\sim 20$ GeV. This effectively places a lower limit of about 
20~GeV on the desired energy of the muons stored in the Neutrino Factory. 
However, a more recent analysis~\cite{nufact06-det} has shown that, with 
more sophisticated selection criteria, high efficiencies ($> 80$\%) can be 
obtained  for neutrino interactions exceeding $\sim 10$ GeV, with 
efficiencies dropping to $\sim 50\%$ by 5 GeV. The new analysis suggests a 
MINOS-like detector could be used at a Neutrino Factory with energy less 
than 20~GeV, but probably not less than 10~GeV. If we wish to consider lower 
energy Neutrino Factories, we will need a finer grained detector that 
enables reliable sign-determination of lower energy muons with good 
efficiency. One way to achieve this would be to use a totally active 
magnetized segmented detector, for example a NOvA-like detector~\cite{Nova} 
within a large magnetic volume. Initial studies~~\cite{nufact06-det} show 
that, for this technology, the muon reconstruction efficiency is expected to 
approach unity for momenta exceeding $\sim 200$~MeV/c, with a charge 
mis-identification lower than $10^{-4}$ ($10^{-3}$) for momenta exceeding 
approximately 400~MeV/c (300~MeV/c).

 Further studies are needed to fully understand the efficiency and 
mis-identification as a function of neutrino energy, but there is hope that 
a Neutrino Factory detector might be designed with good wrong-sign muon 
identification and high efficiency for neutrino energies as low as 1 GeV, or 
perhaps a little lower. Given these recent developments, in our analysis we 
will assume that a massive magnetized detector can be designed to identify 
wrong-sign muons with full efficiency for neutrino interactions above $0.8$ GeV, and zero efficiency below this energy. We will see that the excellent 
physics capability of a low energy Neutrino Factory motivates striving for a 
detector that can achieve this demanding performance.

With a magnetized far detector concept that makes plausible precision 
measurements of neutrino interactions down to about $0.8$ GeV, we are motivated 
to consider Neutrino Factories with stored muon energies of a few GeV. In 
present designs for a 20 GeV Neutrino Factory~\cite{study2a}, there are at least two acceleration stages that accelerate the muons from about 1 GeV to 20 GeV. 
Depending on the design, these accelerators consist either of Recirculating 
Linear Accelerators (RLAs) or Fixed Field Alternating Gradient accelerators 
(FFAGs). A few GeV Neutrino Factory would require only one of these acceleration stages. Note that the RLAs have long straight acceleration 
sections which, if pointing in a suitable direction, could provide a 
neutrino beam with a time-dependent energy that varies from ~200 MeV up to 
the final energy. This might facilitate a tunable-energy Neutrino Factory. 
To illustrate this, Table~\ref{tab:tab1} shows, for $7.5 \times 10^{20}$ 
positive muons (per year) injected into an 1-5.8 GeV RLA, the number of muon 
decays in a given straight section at each intermediate energy. Only $7\%$ 
of the injected muons decay during the acceleration, and hence $93\%$ ($7 
\times 10^{20}$) are available to be injected into a dedicated fixed-energy 
Neutrino Factory. If the Neutrino Factory straight section length is $30\%$ 
of the ring circumference, this would provide an additional $2 \times 
10^{20}$ useful muon decays per year at 5.8~GeV. Note that using the RLA to 
provide a neutrino beam would provide flexibility in choosing the desired 
neutrino energy spectrum. The acceleration cycle could, in principle, be 
varied to keep the muons for as long as desired at any intermediate energy. 
Hence, the $2 \times 10^{20}$ useful muon decays could be redistributed 
amongst the intermediate energies, as needed. However, the flexibility of 
using the RLA to provide a low energy neutrino beam comes at the cost of a 
more complicated design for the accelerator. In particular, the angular 
divergence of the beam within the straight section needs to be small 
compared to the angular spread of the neutrinos generated in muon decay. If 
designs that achieve this prove to be impractical or expensive, flexibility 
could also be achieved by designing the RLA so that the muons could be 
extracted on any given turn and injected into one of several fixed energy 
Neutrino Factory storage rings (note that the cost of the storage rings is 
believed to be small compared to the cost of the RLA).

\begin{table}

\begin{center}

\begin{tabular}{|l|c|c|c|c|c|} \hline

Turn Number                            &  1   &   2  &  3   &   4  &  5  \\

\hline \hline

Initial Energy (GeV)                   & 1.0  & 1.96 & 2.92 & 3.88 & 4.84 \\

Final Energy (GeV)                     & 1.48 & 2.44 & 3.40 & 4.36 & 5.32 \\

$f_{decay}=100m / \gamma c\tau$ (\%)   & 1.30 & 0.73 & 0.51 & 0.39 & 0.32 \\

$N_{decay}$ per year ($\times 10^{18}$)& 9.8  &  5.5 &  3.8 & 2.9  & 2.4\\

\hline

 \end{tabular}

\end{center}

\caption{\it Useful positive muon decays in one straight section of an RLA 
designed to accelerate from 1 GeV to 5.8 GeV in 5 turns. The straight 
section and arc lengths are, respectively, 100m and 30m. The numbers 
tabulated correspond to $7.5 \times 10^{20}$ injected muons, or roughly one 
years operation.}

\label{tab:tab1}

\end{table}

In the following we will show that a low energy Neutrino Factory with a 
fixed energy of 4.12~GeV would provide a sensitive tool for exploring 
neutrino oscillations if $\theta_{13}$ is ''large". This energy would 
require about 4 turns in a single RLA. Note that the sensitivity of a 
Neutrino Factory experiment depends upon the event statistics, and hence 
upon the product of the detector fiducial mass, the length of the data 
taking period, and the number of muons per unit time  decaying in the 
appropriate Neutrino Factory straight section. Initial studies have 
considered, as reasonable, a totally active magnetized detector with a 
fiducial mass of about 20~Kt. Present Neutrino Factory studies suggest that 
it would be reasonable to expect, for a Neutrino Factory with (without) a 
muon cooling channel before the pre-accelerator, about $5 \times 10^{20}$ 
($3 \times 10^{20}$) useful positive muon decays per year and $5 \times 
10^{20}$ ($3 \times 10^{20}$) useful negative muon decays per year in a 
given Neutrino Factory straight section. Hence, a conservative estimate of the sensitivity of a Neutrino Factory experiment 
might be based upon 5 years data taking with  $3 \times 10^{20}$ useful muon 
decays of each sign per year in the storage ring, and a detector fiducial 
mass of 20~Kt, corresponding to  $3 \times 10^{22}$~Kt-decays for each muon 
sign. A more aggressive estimate might be based upon 10 years data taking 
with  $5 \times 10^{20}$ useful muon decays of each sign per year in the 
storage ring, and a detector fiducial mass of 20~Kt, corresponding to  $1 
\times 10^{23}$~Kt-decays for each muon sign.

\section{Numerical analysis: degenerate solutions}
We can ask ourselves whether it is possible to unambiguously determine $\delta$ and $\tetaot$ by measuring the transition probabilities $\nu_e \to \nu_\mu$ and $\bar{\nu}_e \to \bar{\nu}_{\mu}$ at fixed neutrino energy and at just one neutrino factory baseline. The answer is no. 
It has been shown~\cite{nf7b} that, by exploring the full (allowed) range of the $\delta$ and $\tetaot$ parameters, that is, $-180^{\circ}<\delta<180^{\circ}$ and $0^{\circ}<\tetaot<10^{\circ}$, one finds, at fixed neutrino energy and at fixed baseline, the existence of degenerate solutions ($\theta^{'}_{13}$, $\delta^{'}$), that have been labelled in the literature \emph{intrinsic degeneracies}, which give the same oscillation probabilities as the set ($\tetaot$, $\delta$) chosen by nature. More explicitly, if ($\theta_{13}$, $\delta$) are the correct values, the conditions
\begin{equation}
\begin{matrix}
P_{\nu_e \nu_\mu} (\theta^{'}_{13}, \delta^{'}) = P_{\nu_e \nu_\mu}
(\theta_{13}, \delta)~\nonumber \cr 
P_{\bar \nu_e \bar \nu_\mu} (\theta^{'}_{13}, \delta^{'}) = P_{\bar \nu_e
\bar \nu_\mu} (\theta_{13}, \delta)
\end{matrix}
\label{eqn:equalburguet}
\end{equation}
can be generically satisfied by another set ($\theta^{'}_{13}$, $\delta^{'}$). 

Additional solutions might appear from  unresolved degeneracies in two other oscillation parameters:
\begin{enumerate}

\item At the time of the future neutrino factory, the sign of the atmospheric mass difference $\Delta m_{31}^2$ may remain unknown, that is, we would not know if the hierarchy of the neutrino mass spectrum is normal or inverted. In this particular case, $P (\theta^{'}_{13}, \delta^{'}, -\Delta m_{31}^{2}) = P (\theta_{13}, \delta, \Delta m_{31}^2)$.
\item Disappearance experiments only give us information on $\sin^{2} 2 \theta_{23}$: is $\theta_{23}$ in the first octant, or is it in the second one, $(\pi/2 -\theta_{23}$)? . In terms of the probabilities, $P (\theta^{'}_{13}, \delta^{'}, \frac{\pi}{2}-\theta_{23}) = P (\theta_{13}, \delta ,\theta_{23})$.
\end{enumerate}
This problem is known as the problem of degeneracies in the neutrino parameter 
space~\cite{nf7b,nf7c,FL96,MN01,BMWdeg,deg}.
All these ambiguities complicate the experimental determination of $\delta$ and
$\theta_{13}$. Many strategies have been advocated to resolve this issue which in general involve another detector~\cite{MN97,nf7b,silver,BMW02off,SN1,twodetect,SN2,T2kk} or the combination with another experiment~\cite{otherexp1,BMW02,HLW02,MNP03,otherexp,mp2,HMS05,yo,nf9,huber2,lastmine}.

In the present study we show that, if the value of $\theta_{13}$ turns out to be not very small, a \emph{low energy neutrino factory} provides the ideal scenario for the extraction of the unknown parameters as well as for resolving the discrete degeneracies. 
The reason is simple: at distances of $\mathcal{O}(1000)$ km the neutrino oscillation pattern is extremely rich at neutrino energies below $4$ GeV. We have thus explored a single decaying muon energy scenario $E_\mu=4.12$ GeV. 

By exploiting the energy dependence of the signal, it is possible to disentangle $\theta_{13}$ and $\delta$ and to eliminate the additional solutions arising from the discrete ambiguities. We have divided the signal in four energy bins. The energy binning of the signal has been chosen accordingly to the analytical study. The energy range of these four energy bins is $[0, 0.8]$, $[0.8,1.5]$, $[1.5,3.5]$ and $[3.5,4.12]$ GeV.  We will show the physics potential of the chosen energy binning here in the next subsection. The detection efficiencies are considered as perfect ($100\%$) above the first bin, as described in the previous section.

Two possible baselines have been carefully explored: $1280$ Km, the distance from Fermilab to Homestake, and $1480$ Km, the distance from Fermilab to Henderson mine. The results are presented for the two possible scenarios described in the previous section, in order to quantify the benefits of increased exposure times and muon intensities: a conservative scenario of $3\times 10^{22}$~Kton-decays and a more ambitious one with $1 \times 10^{23}$~Kton-decays.  

All numerical results simulated in the next subsections here have been obtained with the exact formulae for the oscillation probabilities. Unless specified otherwise, we take the following central values for the remaining oscillation parameters: $\sin^{2}\theta_{12}=0.29$, $\Delta m^2_{21} =  8 \times 10^{-5}$ eV$^2$,  $|\Delta m^2_{31}| = 2.5 \times 10^{-3}$ eV$^2$ and  $\theta_{23}=40^\circ$ . The $\chi^{2}$ for a fixed baseline $\lambda$ is defined as: 
\begin{equation}
\chi_\lambda^2 = \sum_{i,j} \sum_{p,p'} \; (n^\lambda_{i,p} - N^\lambda_{i,p}) C_{i,p:,j,p'}^{-1} (n^\lambda_{j,p'} - N^\lambda_{j,p'})\,,
\end{equation}
 where $N^\lambda_{i,\pm}$ is the predicted number of muons for
 a certain oscillation hypothesis, $n^\lambda_{i,p}$ are the simulated ``data'' from a Gaussian or Poisson smearing and $C$ is the $2 N_{bin} \times 2 N_{bin}$ covariance matrix, that will contain statistical errors and a $2\%$ overall systematic error. All the contour plots presented in the following in a two parameter space have been performed assuming 2 d.o.f statistics.

\subsection{Optimizing the energy binning}

In this subsection we provide an explanation for the energy binning chosen here, crucial to resolve the additional solutions (degeneracies). As first noticed in Ref.~\cite{otherexp1}, the location of the degeneracies is $E, L$ dependent. For large $\theta_{13}$, the location of the \emph{intrinsic} degeneracies is given by: 
\begin{eqnarray}
\delta^{'}&\simeq&\pi - \delta, \nonumber\\
\theta^{'}_{13} &\simeq & \theta_{13}+ 
\cos \delta \sin 2\theta_{12} \frac{\Delta m^2_{21} L}{4 E} \cot \theta_{23}
\cot \left(\frac{\Delta m^2_{31} L}{4 E}\right)~.
\label{eq:intatm}
\end{eqnarray}
Notice that the shift $\theta^{'}_{13} -\theta_{13}$ depends on the energy and the baseline through the function  $\cot \left(\frac{\Delta m^2_{31} L}{4 E}\right)$. If the function $\cot \left(\frac{\Delta m^2_{31} L}{4 E}\right)$ changes sign from one energy bin to another, the degenerate solutions will appear in different regions of the parameter space. Consequently, the combination of fits to the various energy bins can eliminate the \emph{intrinsic} degeneracies. 

\begin{figure}[h]
\begin{center}
\begin{tabular}{ll}
\includegraphics[width=3in]{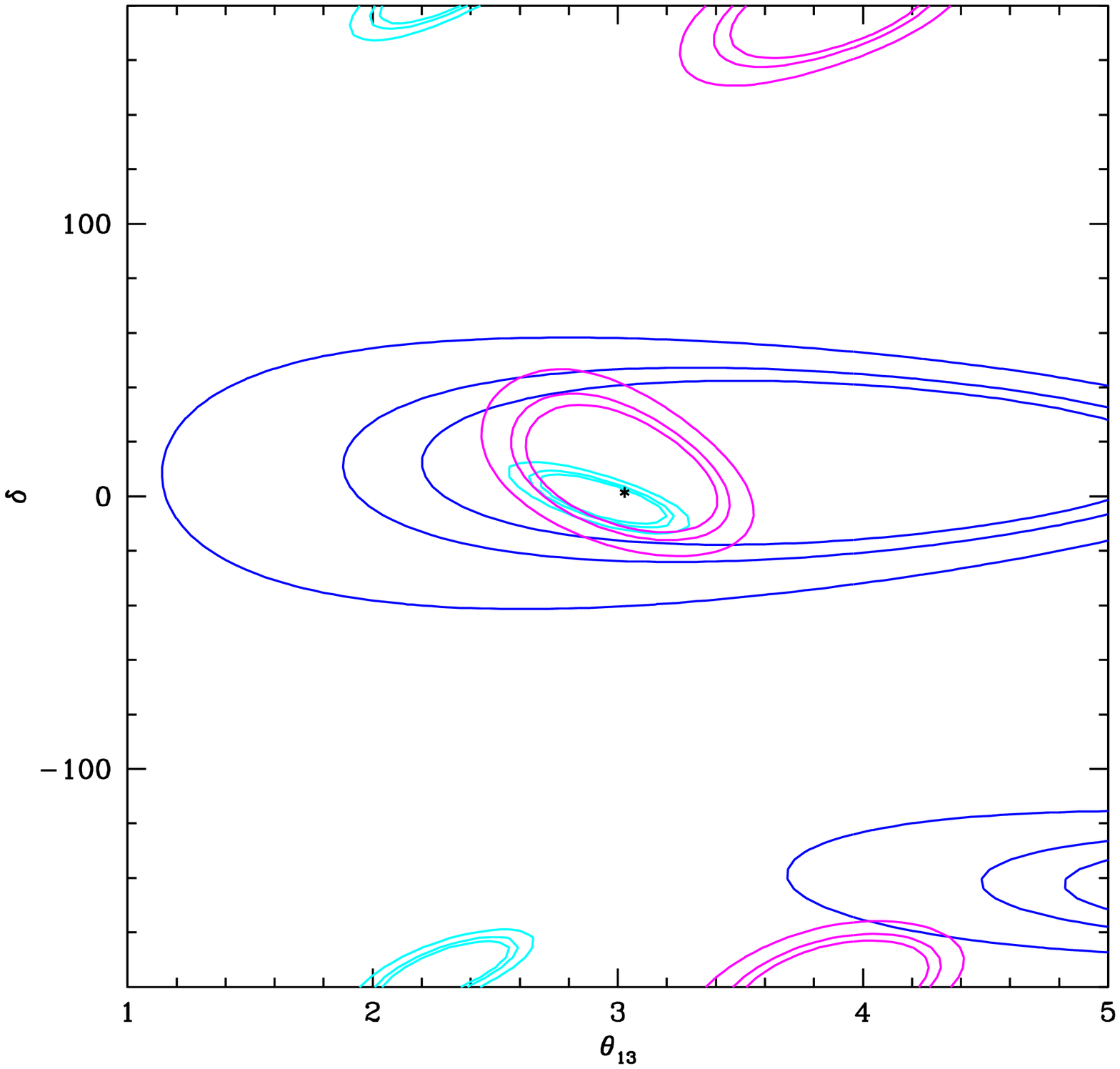}&\hskip 0.cm
\includegraphics[width=3in]{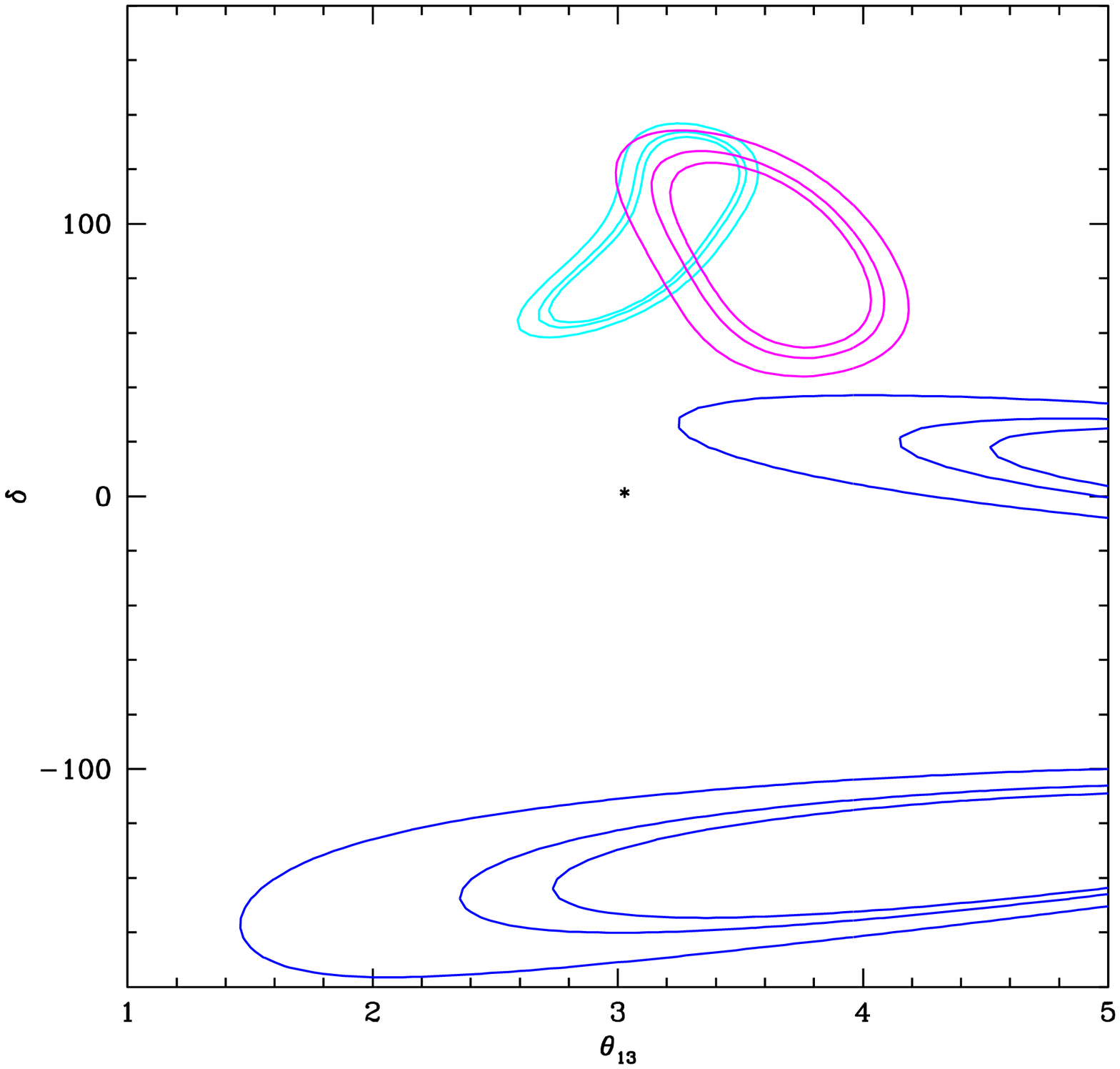}\\
\hskip 2.truecm
{\small (a) Intrinsic degeneracies}            &
\hskip 2.truecm
{\small (b) Sign $\Delta m^2_{31}$ degeneracies}       \\  
\end{tabular}
\end{center}
\caption[]{\textit{$90\%$, $95\%$ and $99\%$ (2 d.o.f) contours resulting from fits to simulated data for $L = 1480$ km, $\theta_{13}=3 ^\circ$, and $\delta=0^\circ$ (position in parameter space denoted by a star).  The blue, cyan and magenta contours represent the results from fits to the second, third and fourth energy bins, respectively. In the left panel, we have neglected the impact of the discrete degeneracies. In the right panel, the fit assumes incorrectly that the sign of the atmospheric mass difference is negative. The atmospheric mixing angle is fixed to $\theta_{23}=40^\circ$.}}
\label{fig:figdeg1}
\end{figure}

To illustrate this, Figure~(\ref{fig:figdeg1})(a) depicts results from fitting the simulated data for each of the energy bins. The simulation is for $L = 1480$ km,  $\theta_{13}=3 ^\circ$, $\delta=0^\circ$, normal mass hierarchy, and $\theta_{23}=40^\circ$. The fit results shown in the figure correspond to the correct hierarchy and $\theta_{23}$ (the impact of including the discrete degeneracies will be discussed later). The $90\%$, $95\%$ and $99\%$ CL contours resulting from the fits are shown for the second energy bin (blue), third energy bin (cyan,) and fourth energy bin(magenta). Notice that in addition to the correct solution, there are also fake solutions for which $\delta^{'}\simeq \pi$, as indicated by Eq.~(\ref{eq:intatm}). However, the fake solutions from the fit to the third energy bin get opposite displacements $\theta^{'}_{13}-\theta_{13}$ from those from the fits to the second and fourth energy bins. The relative displacement (positive or negative) is given by the sign of the trigonometric function  $\cot \left(\frac{\Delta m^2_{31} L}{4 \langle E \rangle}\right)$, where $\langle E \rangle$ is the medium energy for a fixed bin. A combination of fits to the second, third and fourth energy bins will therefore help in resolving the \emph{intrinsic} degeneracies.

A similar exercise to the one described above can be done for the discrete degeneracies. For instance, the wrong-sign$(\deltaunotre)$ additional solutions will be located in different regions of the parameter space for each energy bin, see Fig.~(\ref{fig:figdeg1})(b), and therefore fits to the combination of the energy bins will result in a resolution of the \emph{sign} degeneracies. The location of the fake solutions for the second energy bin is different from those for the third and fourth energy bins, since at lower energies matter effects are less important.

\begin{figure}[h]
\begin{center}
\begin{tabular}{ll}
\includegraphics[width=3in]{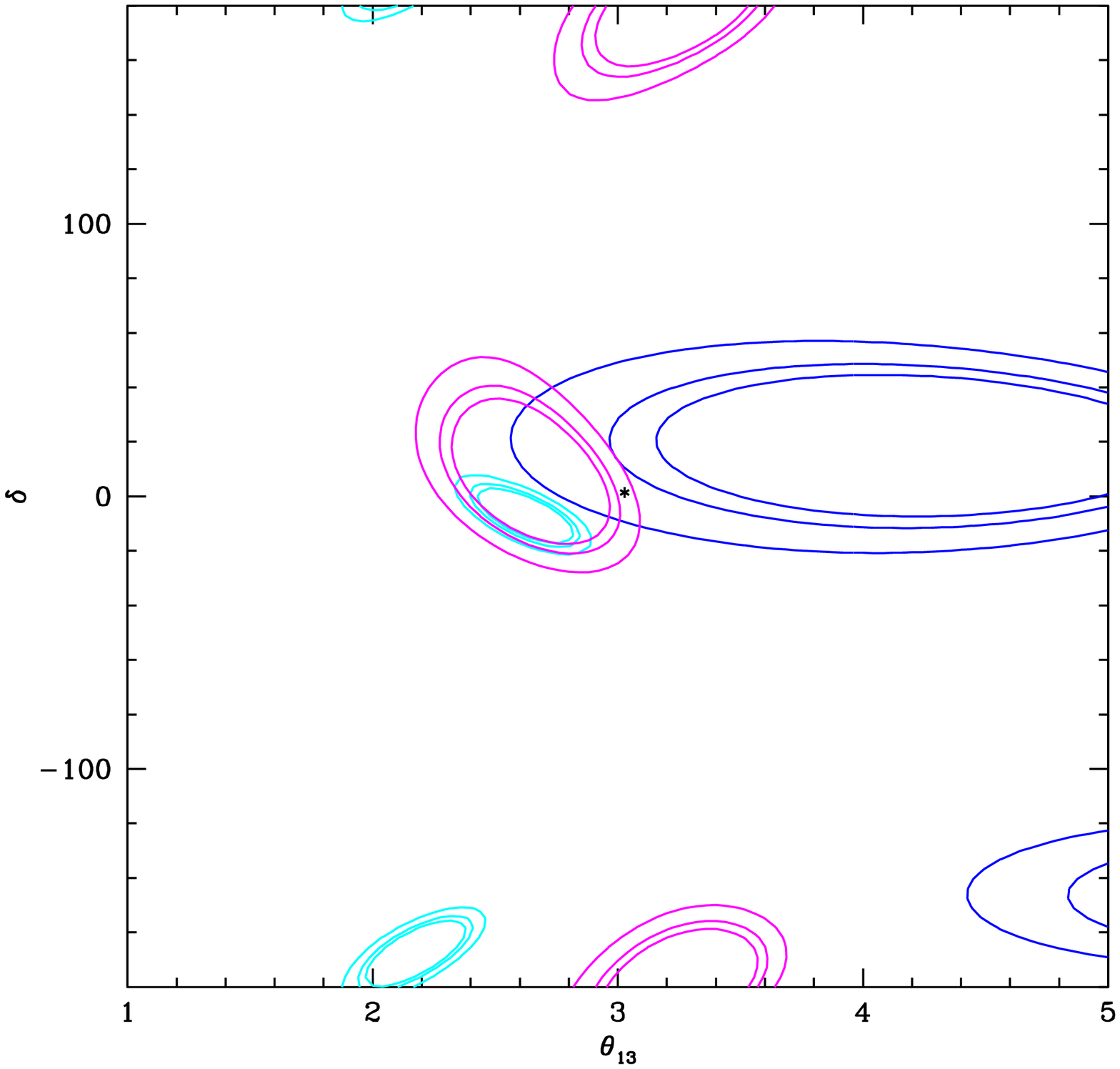}&\hskip 0.cm
\includegraphics[width=3in]{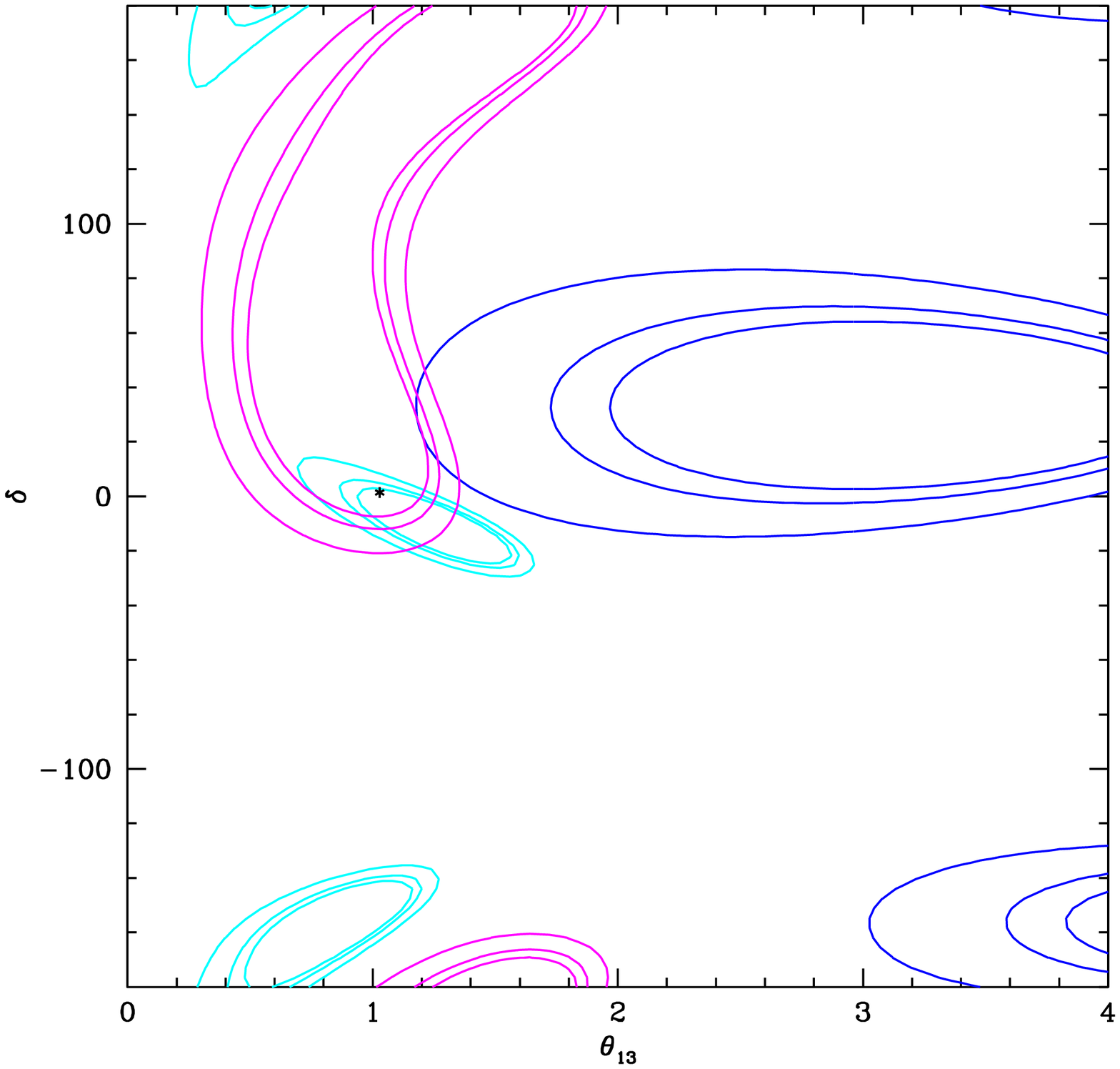}\\
\hskip 2.truecm
{\small (a) $\theta_{13}=3^\circ$}            &
\hskip 2.truecm
{\small (b) $\theta_{13}=1^\circ$}      \\  
\end{tabular}
\end{center}
\caption[]{\textit{(a) $90\%$, $95\%$and $99\%$ (2 d.o.f) contours resulting from the fits to the data at $L = 1480$ km, assuming that the nature solution is $\theta_{13}=3 ^\circ$ and $\delta=0^\circ$ (denoted by a start).  The blue, cyan and magenta contours represent the results with the second, third and fourth energy bin data, respectively. We assume that the true value of the atmospheric mixing angle is $\theta_{23}=40^\circ$ (first octant) but the data is fitted to $\theta_{23}=50^\circ$ (second octant). (b) Same than in (a) but  assuming that the nature solution is $\theta_{13}=1 ^\circ$ and $\delta=0^\circ$ (denoted by a start).}}
\label{fig:figdeg2}
\end{figure}

Resolving the additional $\theta_{23}$-octant degeneracy is, in general, very difficult. As shown in Ref.~\cite{otherexp1}, for large $\theta_{13}$, the location of the $\theta_{23}$ degeneracies is given by: 
\begin{eqnarray}
\sin \delta^{'}&\simeq&\cot \theta_{23}\ \sin \delta, \nonumber\\
\theta^{'}_{13} &\simeq & \tan\theta_{23}\ \theta_{13}+ 
\frac{\sin 2\theta_{12} \frac{\Delta m^2_{21} L}{4 E}}{2 \sin\left(\frac{\Delta m^2_{31} L}{4 E} \right)}\left(\cos\left(\delta-\frac{\Delta m^2_{31} L}{4 E}\right)-\tan\theta_{23} \cos\left(\delta^{'}-\frac{\Delta m^2_{31} L}{4 E}\right)\right).
\label{eq:t23atm}
\end{eqnarray}
This system describes two solutions. For one of them, the $L$ and $E$ dependent terms in Eq.~(\ref{eq:t23atm}) tend to cancel for $\theta_{23} \rightarrow \pi/4$, resulting in $\theta_{13}^{'} =\theta_{13}$ and $\delta^{'}=\delta$ in this limit. The second solution coincides in this limit with the intrinsic degeneracy, Eq.~(\ref{eq:intatm}). 
Notice that no fake solutions are expected  for $|\cot \theta_{23} \;\sin \delta| > 1$. 
Figure~(\ref{fig:figdeg2}) illustrates the equivalent exercise to those performed above for the \emph{intrinsic} and for the wrong-sign$(\deltaunotre)$ degeneracies. The simulated data is for the $\theta_{23}$ in the first octant, i.e. $\theta_{23}=40^\circ$, while the fit is performed incorrectly assuming the second octant, i.e. $\theta_{23}=50^\circ$.  Notice from the results depicted in Fig.~(\ref{fig:figdeg2})(a) that there are two sets of degenerate solutions, as previously discussed: those which resemble the correct values and those which are related to the intrinsic solution. While the location of the former is $E, L$ independent, the location of the latter will depend on $E, L$, and therefore the combined fits to the various energy bins will help in resolving these degeneracies. Note that the degeneracies which are closer to the correct values are extremely difficult to resolve.  The information in the second bin is crucial: if the detector efficiency in the second bin ($[0.8,1.5]$ GeV) is sufficiently high (we are assuming $100\%$), the combination of fits to the various energy bins will resolve the additional solutions related to the wrong choice of the atmospheric mixing angle octant. Since the second bin is the lower energy bin, it turns out that for these lower energies the solar term, see Eq.~(\ref{eq:probappr}), is the dominant one for $\theta_{13} \le 3^\circ$, larger than both the atmospheric and the interference terms. The solar term goes as $\cos^2 \theta_{23}$, while the atmospheric term goes as $\sin^2 \theta_{23}$, therefore, exploiting this low energy bin is crucial to resolve the $\theta_{23}$ degeneracy.  The resulting fit, after the combination of the data in the three energy bins considered here, is degeneracy free down to very small values of $\theta_{13}\simeq 1^\circ$ (see Fig.~(\ref{fig:figdeg2})(b)), since the additional solutions from fits to the second bin lie on different locations in parameter space than those for the third and fourth energy bins.

\subsection{Exploring the disappearance channel}

\begin{figure}[h]
\begin{center}
\begin{tabular}{ll}
\includegraphics[width=3in]{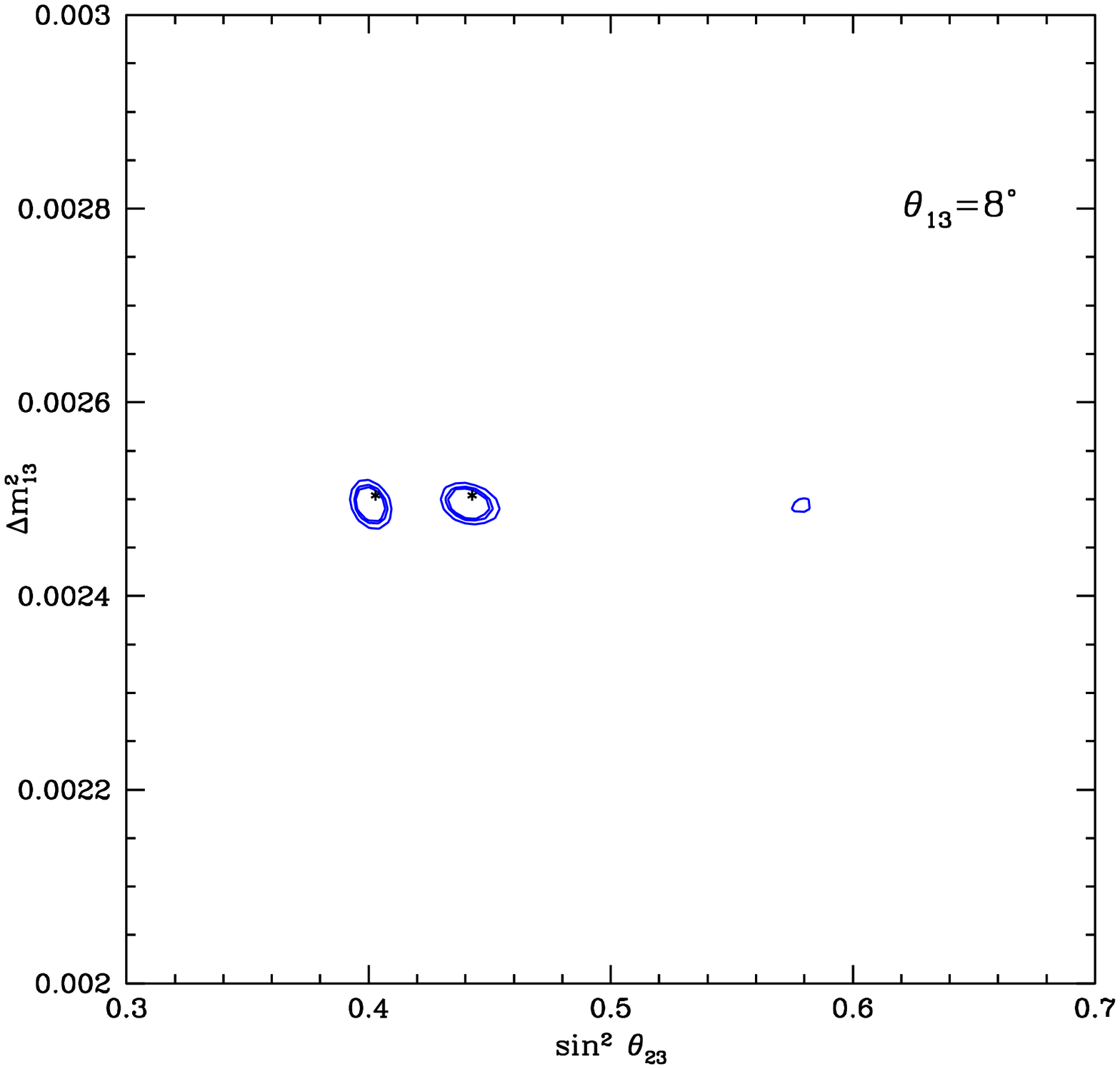}&\hskip 0.cm
\includegraphics[width=3in]{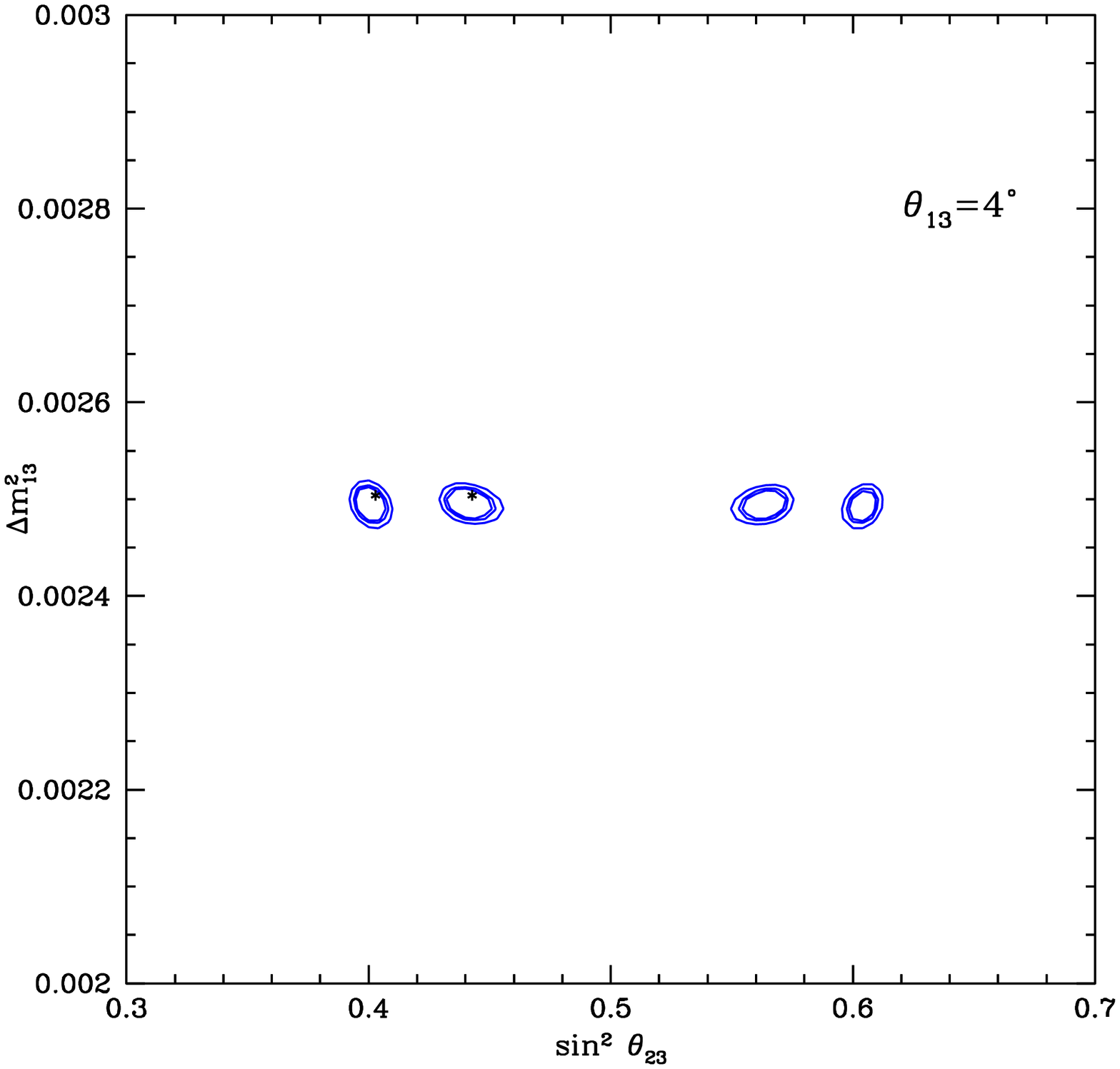}\\
\hskip 2.truecm
{\small (a) $\theta_{13}=8^\circ$}            &
\hskip 2.truecm
{\small (b) $\theta_{13}=4^\circ$}       \\  
\end{tabular}
\end{center}
\caption{\textit{$90\%$, $95\%$ and $99\%$ (2 d.o.f)  CL contours resulting from the fits at $L = 1480$ km assuming two central values for $\sin^2 \theta_{23}=0.4$ and $0.44$ and  $\Delta m^2_{31}=2.5 \times 10^{-3}$ eV~$^2$. In the right (left) panel, $\theta_{13}=8^\circ$ ($4^\circ$). A larger value of $\theta_{13}$ introduces a larger asymmetry and the four-fold degeneracy in the atmospheric neutrino parameters is solved. 
The statistics considered for both simulations corresponds to $3\times 10^{22}$~Kton-decays.  Only disappearance data have been used to perform these plots.} }   
\label{fig:dis}
\end{figure}

We explore here the measurement of the atmospheric neutrino oscillation parameters, $\theta_{23}$ and $\Delta m^2_{31}$ making use of the $\nu_\mu$ disappearance channel. We study as well the impact of this channel regarding the $\theta_{23}$-octant degeneracy. The disappearance channel at the neutrino factory has been already considered in the literature~\cite{nf4,dis} and it has been widely and carefully explored in Ref.~\cite{Stef} . The vacuum $\nu_\mu \to \nu_\mu$ oscillation probability expanded to the second order in the small parameters $\theta_{13}$ and $(\Delta_{12}L/E)$ reads~\cite{Akhmedov:2004ny}
\bea
P(\nu_\mu \to \nu_\mu) & = & 1-  \left [ \sin^2 2 \theta_{23} -s^2_{23} \sin^2 2 \theta_{13} \cos
    2\theta_{23} \right ]\, \sin^2\left(\frac{\Delta_{23} L}{2}\right) \cr
& - & \left(\frac{\Delta_{12} L}{2}\right) [s^2_{12} \sin^2 2 \theta_{23} + \tilde{J} 
s^2_{23} \cos \delta] \, \sin(\Delta_{23} L) \cr
& - & \left(\frac{\Delta_{12} L}{2}\right)^2 [c^4_{23} \sin^2 2\theta_{12}+
s^2_{12} \sin^2 2\theta_{23} \cos(\Delta_{23} L)]~,
\label{eq:probdismu}
\eea
where $\tilde{J}=\cos \theta_{13} \sin 2\theta_{12}\sin 2\theta_{13}\sin 2\theta_{23}$ and $\Delta_{23}=\Delta m^2_{32}/2 E$, $\Delta_{12}=\Delta m^2_{21}/2 E$.
The first term in the first parenthesis is the dominant one and is symmetric under $\theta_{23} \to \pi/2-\theta_{23}$. However, when a rather large non-vanishing $\theta_{13}$ is switched on, a $\theta_{23}$-asymmetry appears in Eq.~(\ref{eq:probdismu}) and the octant in which $\theta_{23}$ lies can be extracted from disappearance data, as will be shown in our numerical results.  
We assume here the same detection efficiencies\footnote{We believe this is  conservative since less aggressive cuts are required to reduce backgrounds for the disappearance channel than those required for the appearance channel.} and energy binning than those which will be considered for the golden $\nu_e \to \nu_\mu$ transition. 
A global $2 \%$ systematic error has been included in the $\chi^2$ fits to the atmospheric neutrino parameters. 
We present our results in the ($\sin^2\theta_{23},\Delta m^2_{31}$) plane in Figs.~(\ref{fig:dis}) for  two simulated values of $\theta_{13}$, and two simulated values for $\sin^2 \theta_{23}$: $\sin^2 \theta_{23}=0.4$ and $\sin^2 \theta_{23}=0.44$. The detector is located at the Henderson mine at a baseline of $L=1480$ km (similar results are obtained for the Homestake detector location).  The CP violating phase $\delta$ has been set to zero.  
Notice that this channel is able to reduce atmospheric parameter uncertainties to an unprecedented level: the resolution in $\sin^2\theta_{23}$ is astonishing, maximal mixing can be excluded at $99\%$ CL if $\sin^2 \theta_{23}<0.48$ ($\theta_{23}<43.8^\circ$), independently of the value of $\theta_{13}$. In addition, for a relatively \emph{large} value of $\theta_{13}>8^\circ$, the $\theta_{23}$-octant degeneracy will not be present at the $99\%$ CL for $\sin^2 \theta_{23}<0.44$ ($\theta_{23} < 41.5^\circ$), if $\theta_{13}$ is treated as a fixed parameter. These results have been obtained for the more conservative neutrino factory scenario described above, a scenario with $3 \times 10^{22}$~Kton-decays for each muon sign. Since the statistics and the size of the expected signal are both large in disappearance measurements, the error on the parameters will be dominated by the systematic error and 
a more ambitious scenario with higher statistics (with $1 \times 10^{24}$~Kton-decays) will not improve much these results.

\subsection{Simultaneous fits to $\theta_{13}$ and $\delta$}
\begin{figure}[h]
\begin{center}
\begin{tabular}{ll}
\includegraphics[width=3in]{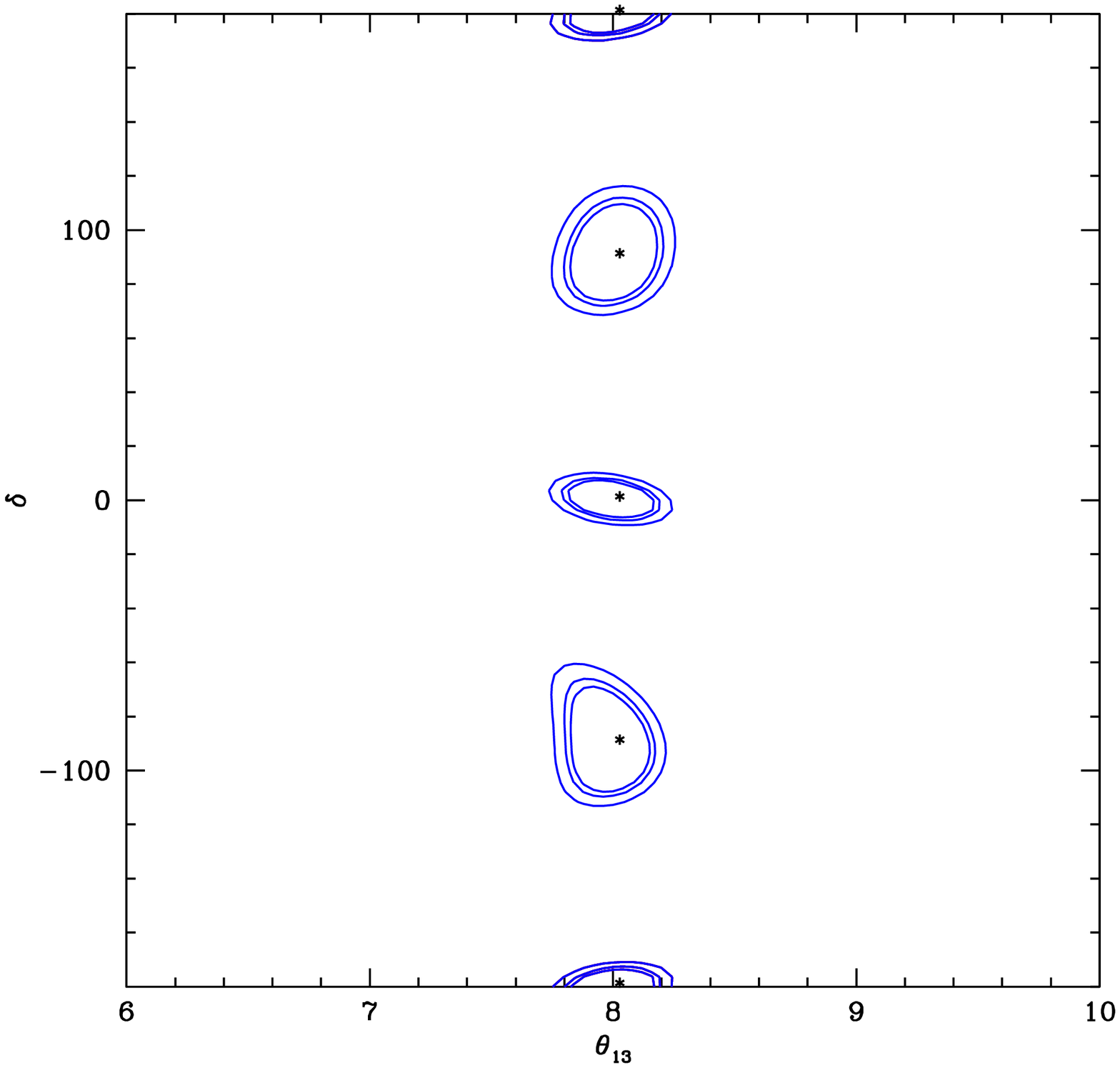}&\hskip 0.cm
\includegraphics[width=3in]{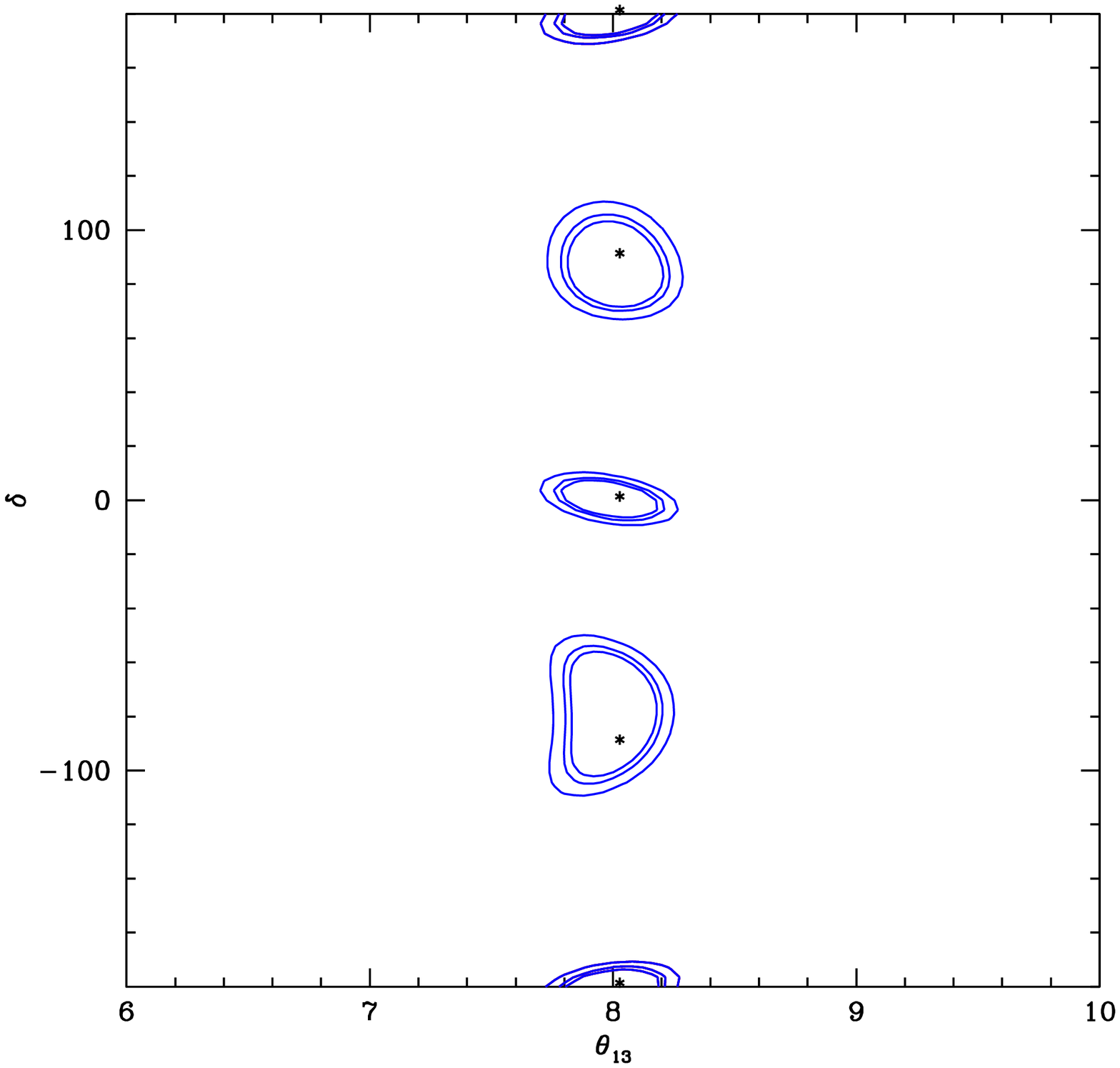}\\
\hskip 2.truecm
{\small (a) Homestake, $L=1280$ km.}            &
\hskip 2.truecm
{\small (b) Henderson, $L=1480$ km}       \\  
\end{tabular}
\end{center}
\caption[]{\textit{$90\%$, $95\%$ and $99\%$ (2 d.o.f) CL contours resulting from the fits at $L = 1280$ km (left panel) and  $L = 1480$ km (right panel) assuming four central values for $\delta=0^{\circ}$, $90^{\circ}$, $-90^{\circ}$ and $180^{\circ}$ and  $\tetaot=8^\circ$. The statistics considered for both simulations corresponds to $3\times 10^{22}$~Kton-decays.}}
\label{fig:fig8}
\end{figure}

In this subsection we exploit the \emph{golden channel}, i.e. the $\nu_e (\bar{\nu}_e) \to \nu_\mu (\bar{\nu}_\mu)$ transitions to extract the unknown parameters $\theta_{13}$ and $\delta$.
We start exploring the more conservative neutrino factory scenario with $3 \times 10^{22}$~Kton-decays. We present in Figs.~(\ref{fig:fig8}) the $90\%$, $95\%$ and $99\%$ CL contours for a fit to the simulated data from a future low energy neutrino factory with the detector located at Homestake, at a baseline $L=1280$ km (left panel) and at Henderson, at a baseline $L=1480$ km (right panel). The ``true'' parameter values that we have chosen for these examples are depicted in the figures with a star: we have explored four different values of $\delta=0^{\circ}$, $90^{\circ}$, $-90^{\circ}$ and $180^{\circ}$ and  $\tetaot=8^\circ$. The simulations are for the normal mass hierarchy and $\theta_{23}$ in the first octant ($\sin^2 \theta_{23} = 0.41$ which corresponds to $\theta_{23}=40^\circ$).
Our analysis includes the study of the discrete degeneracies. That is, we have fitted the data assuming both the wrong hierarchy and the wrong choice for the $\theta_{23}$ octant (i.e. negative hierarchy and $\sin^2 \theta_{23}=0.59$, which corresponds to $\theta_{23}=50^\circ$) and the additional solutions (if present) will be shown in red and in cyan, respectively. 
Notice that in Figs.~(\ref{fig:fig8}) the sign ambiguity is solved at the $99\%$ CL. The additional solutions associated to the wrong choice of the $\theta_{23}$ octant are not present at the same CL due to the information extracted from the disappearance channel.

\begin{figure}[h]
\begin{center}
\begin{tabular}{ll}
\includegraphics[width=3in]{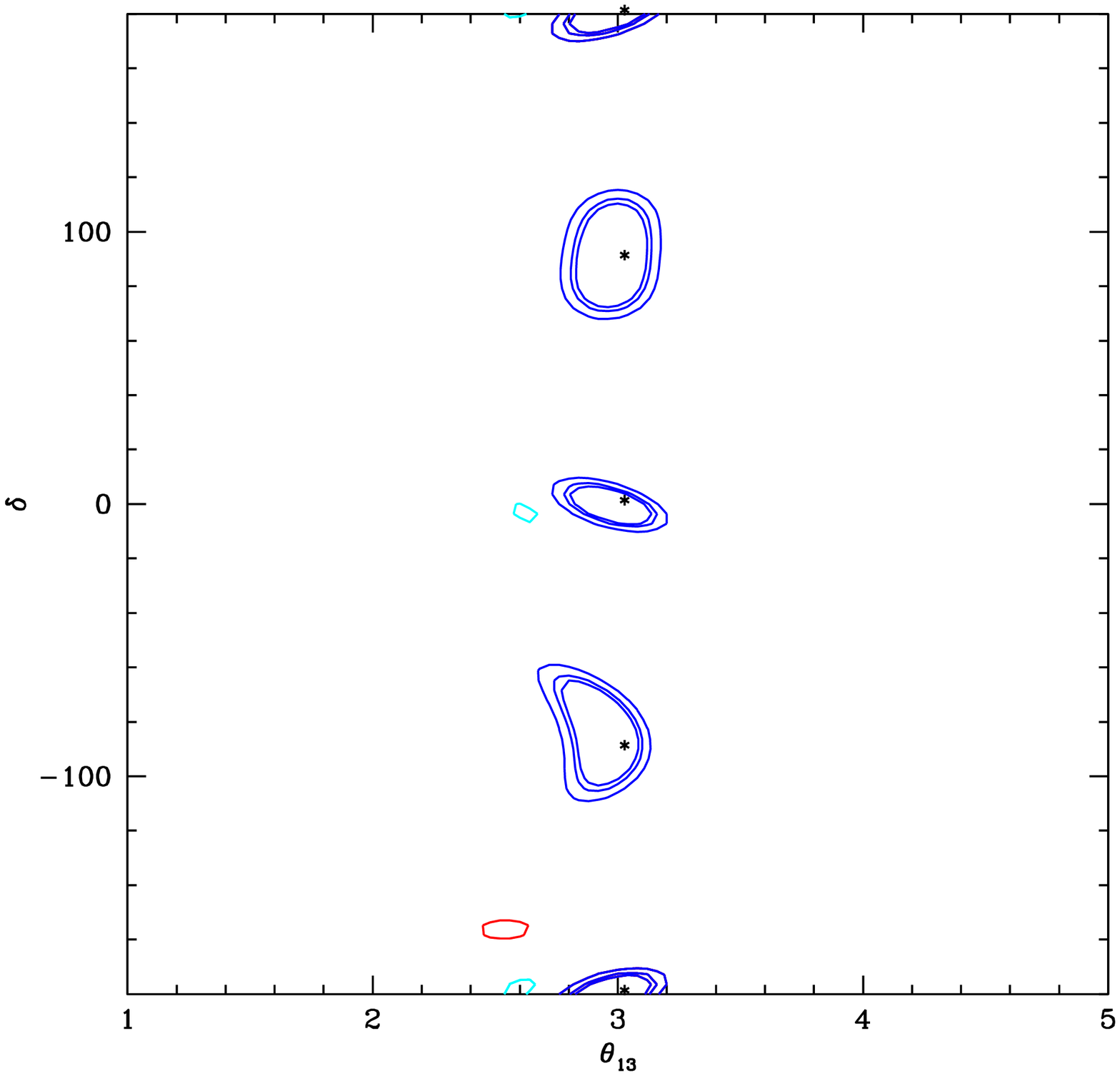}&\hskip 0.cm
\includegraphics[width=3in]{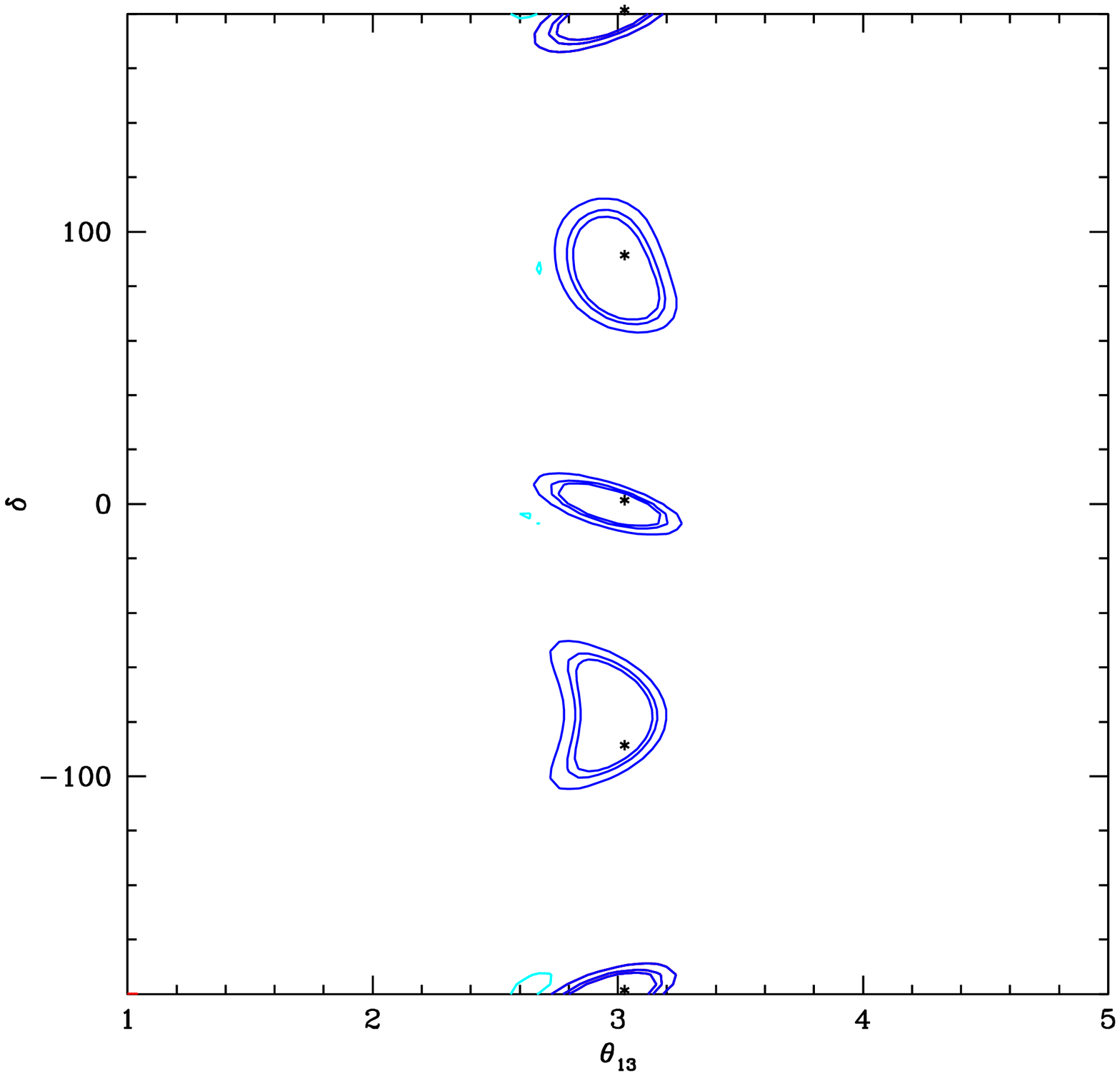}\\
\hskip 2.truecm
{\small (a) Homestake, $L=1280$ km.}            &
\hskip 2.truecm
{\small (b) Henderson, $L=1480$ km}       \\  
\end{tabular}
\end{center}
\caption[]{\textit{ The same as Figs.~(\ref{fig:fig8}) but for $\tetaot=3^\circ$. The additional solution (only at the $99\%$ CL) associated to the wrong choice of the neutrino mass ordering is depicted in red. The additional solutions (only at the $99\%$ CL) arising from the wrong choice of the $\theta_{23}$ octant are depicted in cyan. The statistics considered for both simulations corresponds to $3\times 10^{22}$~Kton-decays.}}
\label{fig:fig3}
\end{figure}
Similar results are obtained for smaller values of $\theta_{13}$, see Figs.~(\ref{fig:fig3}). Notice that the performance of the \emph{low energy neutrino factory} is unique: 
the sign($\Delta m_{31}^2$) can be determined at the $99\%$ if $\theta_{13}>2^\circ$ independent of the value of the CP phase $\delta$. Regarding the $\theta_{23}$-octant ambiguity, it can be removed at the $99\%$ CL down roughly to $\theta_{13}> 1^\circ$ for a nature's choice of $\sin^2 \theta_{23}=0.41$, independent of the value of $\delta$, provided that the conservative estimate of $3 \times 10^{22}$~Kt-decays for each muon sign are feasible. The $\theta_{23}$ octant degeneracy is solved with the information contained in the second energy bin data, which is sensitive to the solar term, as mentioned before. 

\begin{figure}[h]
\begin{center}
\begin{tabular}{ll}
\includegraphics[width=3in]{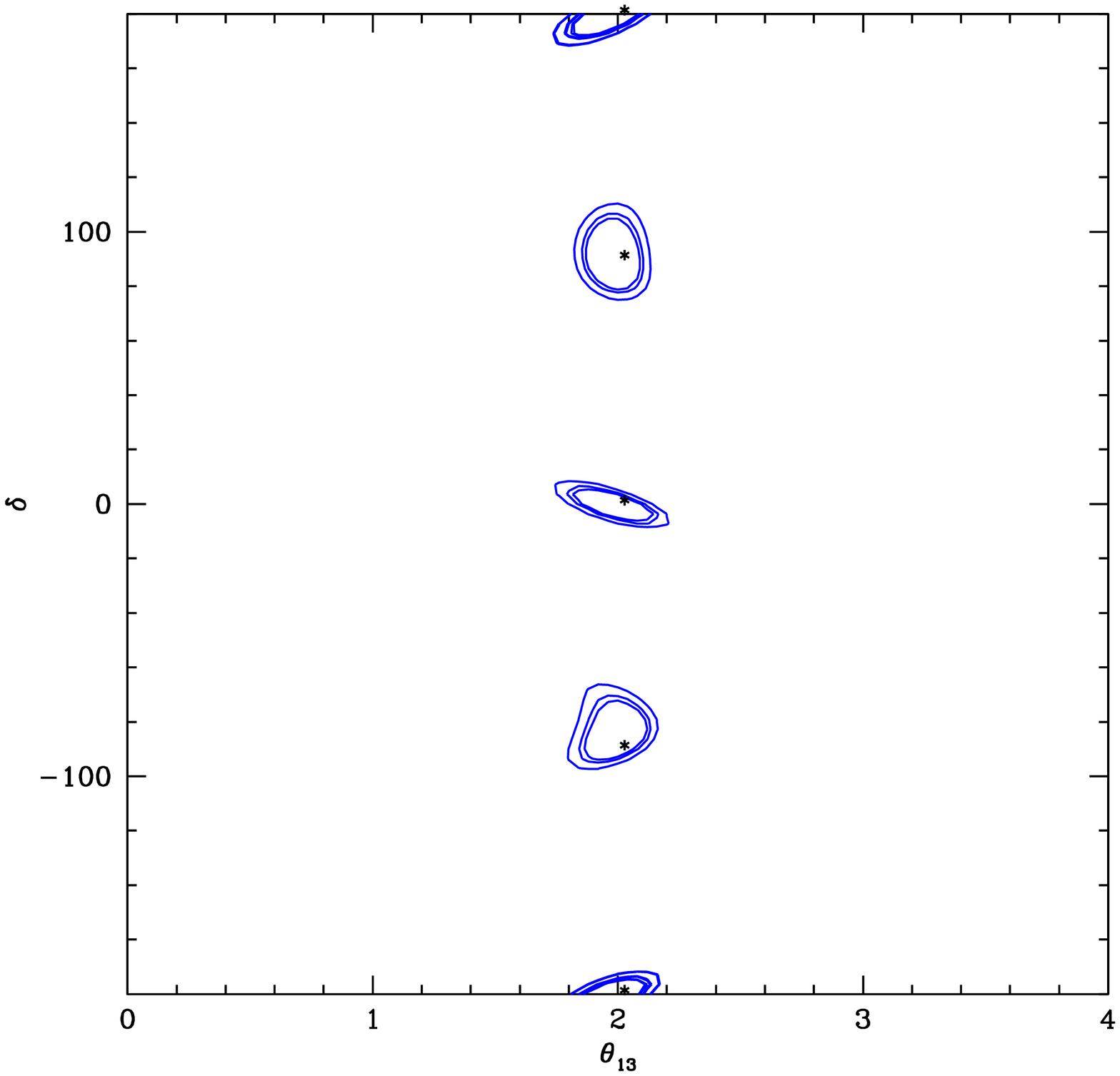}&\hskip 0.cm
\includegraphics[width=3in]{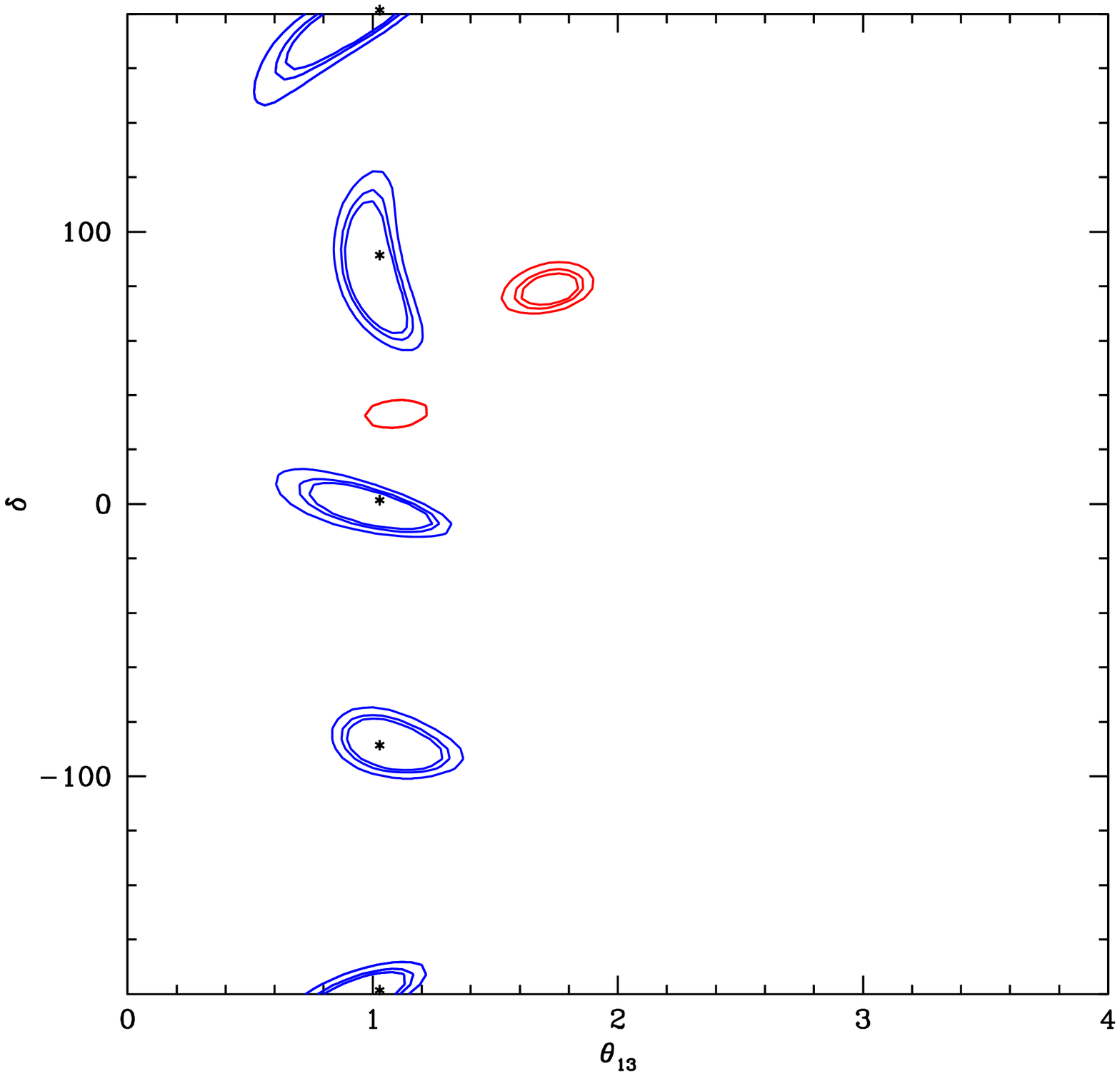}\\
\hskip 2.truecm
{\small (a) $\theta_{13}=2^\circ$}            &
\hskip 2.truecm
{\small (b)  $\theta_{13}=1^\circ$}      \\  
\end{tabular}
\end{center}
\caption[]{\textit{
$90\%$, $95\%$ and $99\%$ (2 d.o.f) CL contours resulting from the fits at $L = 1480$ km assuming four central values for $\delta=0^{\circ}$, $90^{\circ}$, $-90^{\circ}$ and $180^{\circ}$ and  $\tetaot=2^\circ$ in the left panel ($\tetaot=1^\circ$ in the right panel). The additional solutions associated to the wrong choice of the neutrino mass ordering are depicted in red. 
The statistics considered for both simulations corresponds to $1\times 10^{23}$~Kton-decays.}} 
\label{fig:agr}
\end{figure}

Since the results are very similar for the two baselines explored here, the physics reach with the more aggressive estimate of $1 \times 10^{23}$~Kt-decays for each muon sign is illustrated for only one baseline, $L=1480$ km (Henderson mine site) and for  smaller values of $\theta_{13}$. Figure~(\ref{fig:agr}) shows fit results for two simulated values of $\theta_{13}$ ($2^\circ$ and $1^\circ$). The mass hierarchy can be determined at the $99\%$ CL if $\theta_{13}>1^\circ$ independent of the value of the CP phase $\delta$. In addition, for our example with $\sin^2 \theta_{23}=0.41$, the $\theta_{23}$ octant ambiguity can be resolved at $99\%$CL for all values of the CP phase $\delta$ provided $\theta_{13} > 0.6^\circ$.

We summarize the reach of the low energy neutrino factory with two exclusion plots which illustrate the performance of the experiment explored here. We have taken into account the impact of both the intrinsic and discrete degeneracies to depict the excluded regions.
For both exclusion plots we have assumed that the detector is located at the Henderson mine at $L=1480$ km. The results for the closer baseline ($L=1280$ km) are very similar.

\begin{figure}[h]
\begin{center}
\includegraphics[width=3.5in]{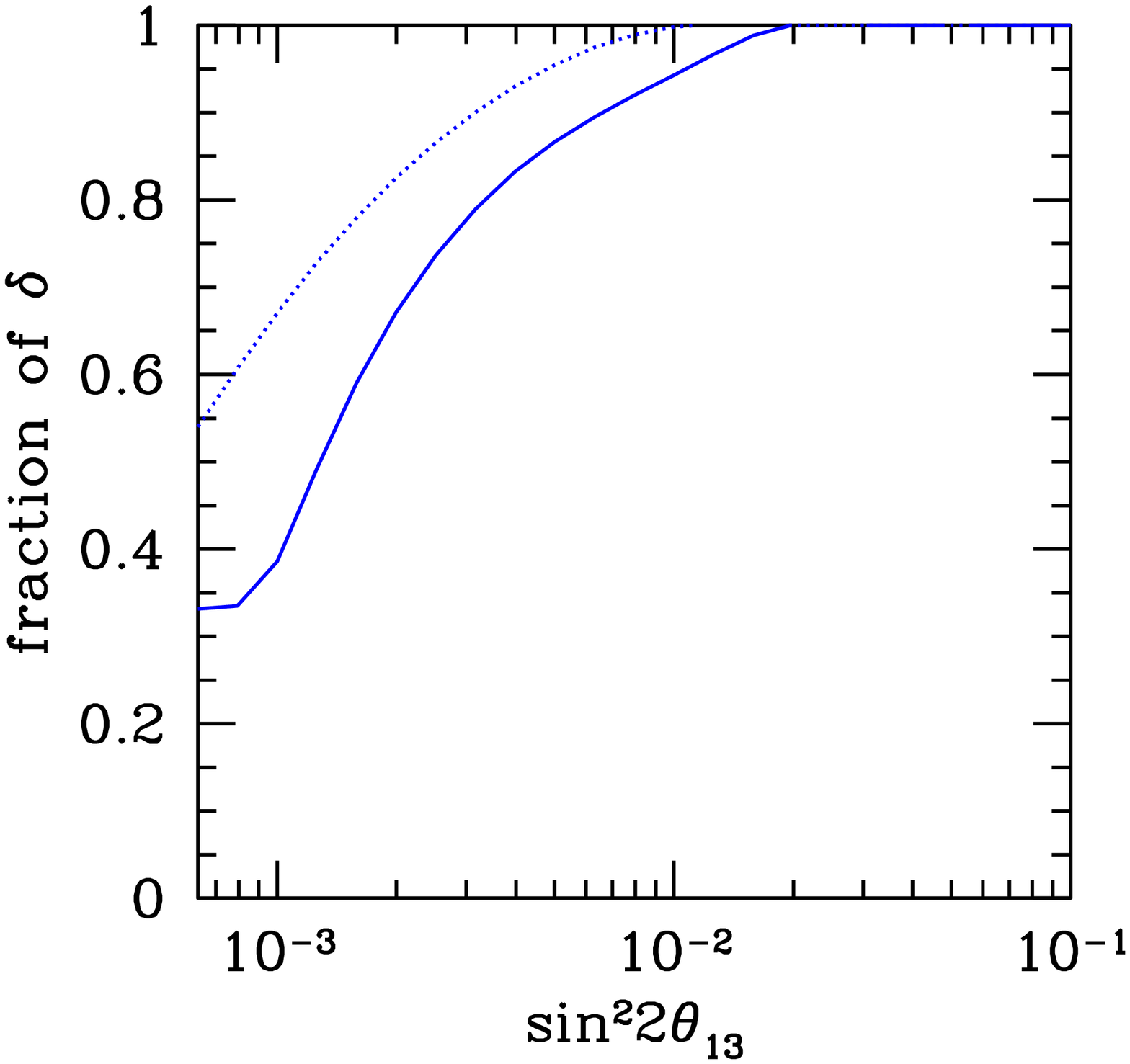}
\end{center}
\caption[]{\textit{$95\%$ CL hierarchy resolution (2 d.o.f) assuming that the far detector is located at a distance of $1480$ km at the Henderson mine. The solid (dotted) curves depict the results assuming  $3\times 10^{22}$~Kton-decays ($1\times 10^{23}$~Kton-decays).}}
\label{fig:hier}
\end{figure}

Figure~(\ref{fig:hier}) depicts the region in the $\sin^2 2 \theta_{13}$, fraction of $\delta$ plane for which the hierarchy can be resolved  at the $95\%$ CL assuming 2 d.o.f statistics, for both scenarios, the more conservative one, in which the exposure is  $3\times 10^{22}$~Kton-decays, and the more aggressive scenario in which the exposure is  $1\times 10^{23}$~Kton-decays. 
Notice that the hierarchy could be determined in both scenarios if $\sin^2 2 \theta_{13}>0.01$ (i.e. $\theta_{13}> 3^\circ$) regardless of the value of the CP violating phase $\delta$.

\begin{figure}[h]
\begin{center}
\includegraphics[width=3.5in]{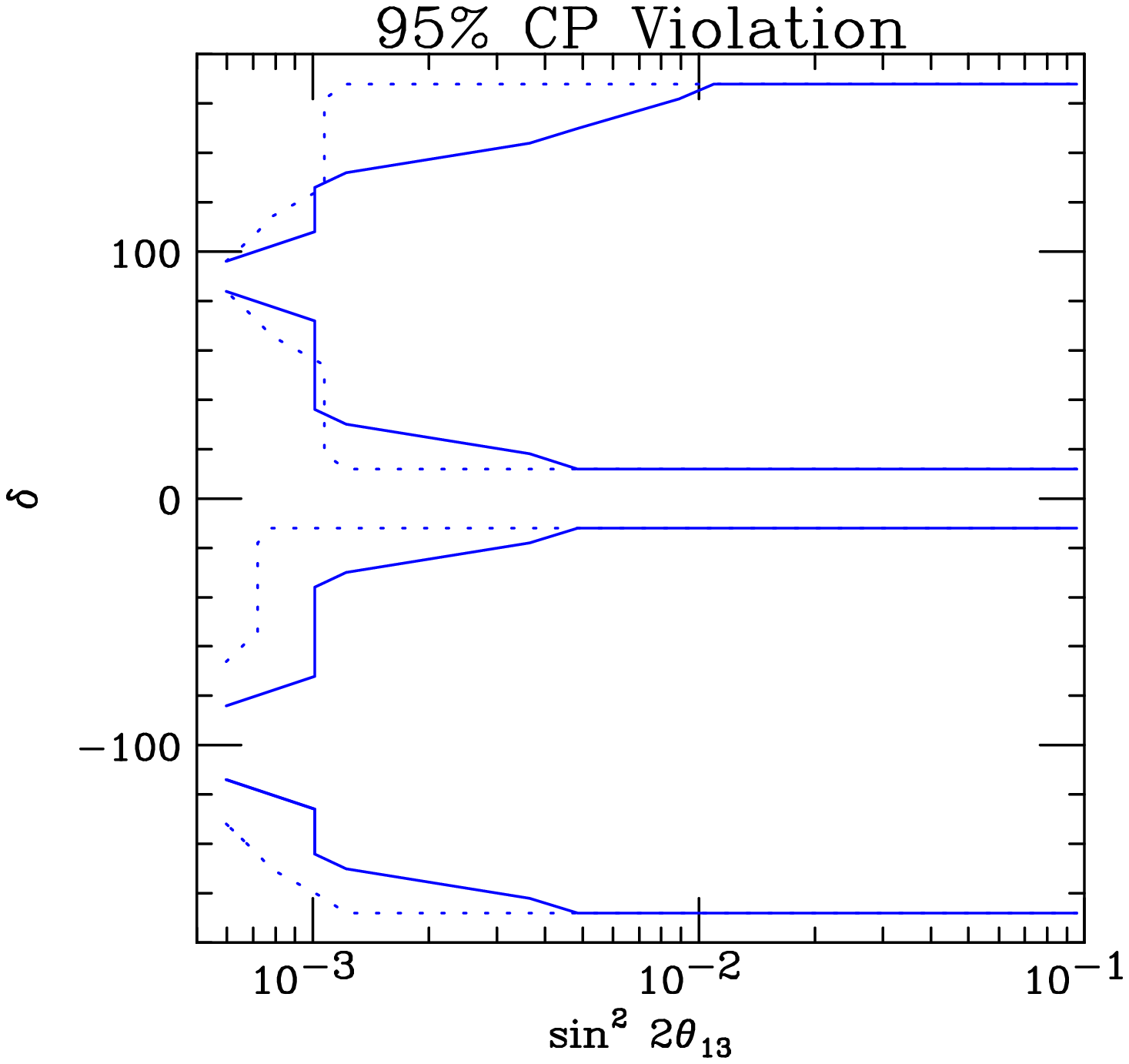}
\end{center}
\caption[]{\textit{$95\%$ CL CP Violation  extraction (2 d.o.f) assuming that the far detector is located at a distance of $1480$ km at the Henderson mine. The solid (dotted) curves depict the results assuming  $3\times 10^{22}$~Kton-decays ($1\times 10^{23}$~Kton-decays).}}
\label{fig:cp}
\end{figure}

Figure~(\ref{fig:cp}) depicts the region in the ($\sin^2 2 \theta_{13}$, $\delta$) plane for which a given (non-zero) value of the CP violating phase can be distinguished  at the $95\%$ CL from the CP conserving case, i.e. $\delta =0, \pm 180^\circ$ (assuming 2 d.o.f statistics). This exercise is illustrated for the two scenarios considered in this study, the more conservative one, in which the exposure is  $3\times 10^{22}$~Kton-decays, and the more agressive scenario in which the exposure is  $1\times 10^{23}$~Kton-decays.  Notice that the CP violating phase $\delta$ could be measured with a $95\%$ CL error lower than $20^\circ$ in both scenarios if $\sin^2 2 \theta_{13}>0.01$ (i.e. $\theta_{13}> 3^\circ$), reaching an unprecedent precision for larger values of $\theta_{13}$. For smaller values, $0.001<\sin^2 2 \theta_{13}<0.01$, and in the more conservative scenario, the presence of the sign-degeneracy compromises the extraction of the CP violating phase $\delta$. On the other hand, in the more aggressive scenario, a CP violating effect could be established at the $95\%$ CL if $\sin^2 2 \theta_{13} \ge 0.001$ for $20^\circ <\delta <160^\circ$ ($-160^\circ <\delta <-20^\circ$).

\section{Conclusions}

We have shown here the enormous physics reach of a novel neutrino factory concept, a \emph{low energy neutrino factory}, in which the stored muons have an energy of $4.12$ GeV. 

We have exploited both the disappearance ($\nu_\mu \to \nu_\mu$)  and the \emph{golden} ($\nu_e \to \nu_\mu$) channels by measuring the ``right-sign'' and the ``wrong-sign'' muons at a two possible baselines: $1280$ Km, the distance from Fermilab to Homestake, and $1480$ Km, the distance from Fermilab to Henderson mine.  The results presented here can be easily generalized to other baselines in the 1200--1500~km range.

We illustrate the results of the analysis of the energy binned signal for a facility with (a) $3 \times 10^{22}$~Kt-decays for each muon sign and (b) $1 \times 10^{23}$~Kt-decays for each muon sign. The novel setup presented here could extract the $\theta_{13}$ angle, the neutrino mass hierarchy and the leptonic CP violating phase $\delta$ with unprecedented precision.

The unique performance of the \emph{low energy neutrino factory} (when compared to the common $20-50$ GeV neutrino factory) is due to the rich neutrino oscillation pattern at energies between $1$ and $4$ GeV at baselines $\mathcal{O}(1000)$ km. Recent studies have shown that it could be possible a Neutrino Factory detector with good wrong-sign muon identification and high efficiency for neutrino energies as low as 1 GeV, or perhaps a little lower. Therefore, to evaluate the physics potential of a low energy neutrino factory, we have assumed $100\%$ efficiency above a threshold energy of 0.8 GeV, and zero efficiency below this threshold. This na\"{i}ve model for the detector performance will need to be updated once further work has been done to better understand the expected detector energy dependent efficiency.

With this caveat we find that maximal atmospheric neutrino mixing can be excluded at $99\%$ CL if $\sin^2 \theta_{23}<0.48$ ($\theta_{23}<43.8^\circ$).  If the atmospheric mixing angle is not maximal, for a nature's choice of $\sin^2 \theta_{23}=0.4$, the octant in which $\theta_{23}$ lies could be extracted at the $99\%$ CL if $\theta_{13}> 1^\circ$ ($\theta_{13}> 0.6^\circ$) with an exposure of $3 \times 10^{22}$~Kt-decays ($1 \times 10^{23}$~Kt-decays) for each muon sign, independently of the value of the CP violating phase $\delta$.  The neutrino mass hierarchy could be determined at the $95\%$ CL, and the CP violating phase $\delta$ could be measured with a $95\%$ CL error lower than $20^\circ$, if $\sin^2 2 \theta_{13}>0.01$ (i.e. $\theta_{13}> 3^\circ$) assuming the more conservative exposure scenario. All the sensitivities quoted here are computed assuming the $2$ d.o.f statistical approach, and in our analysis we have included statistical and a $2\%$ overall systematic error.

In summary, the \emph{low energy neutrino factory} scenario could provide the ideal laboratory for precision lepton physics if the mixing angle $\theta_{13}>2^\circ$.

\vspace{1cm}
\section*{Acknowledgments} 

We thank Stephen Parke for stimulating ideas. O.~M. would like to thank D.~Meloni for comments on the manuscript. This work was supported in part by the European Programme ``The Quest for
Unification''  contract MRTN-CT-2004-503369, and by the Fermi National Accelerator Laboratory, which is operated by the Fermi Research Association, under contract No. DE-AC02-76CH03000 with the U.S. Department of Energy.


\begin{thebibliography}{99}
\bibitem{sol}
  B.~T.~Cleveland {\it et al.},
  Astrophys.\ J.\  {\bf 496}, 505 (1998);
  Y.~Fukuda {\em et al.} [Kamiokande Collaboration],
  Phys.\ Rev.\ Lett.\  {\bf 77}, 1683 (1996);
  J.~N.~Abdurashitov {\it et al.}  [SAGE Collaboration],
  J.\ Exp.\ Theor.\ Phys.\  {\bf 95}, 181 (2002);
  W.~Hampel {\it et al.}  [GALLEX Collaboration],
  Phys.\ Lett.\ B {\bf 447}, 127 (1999);
  T.~A.~Kirsten  [GNO Collaboration],
  Nucl.\ Phys.\ Proc.\ Suppl.\  {\bf 118}, 33 (2003).
\bibitem{SKsolar}
  S.~Fukuda {\it et al.}  [Super-Kamiokande Collaboration],
  Phys.\ Lett.\ B {\bf 539}, 179 (2002).
\bibitem{SNO1}
  Q.~R.~Ahmad {\it et al.}  [SNO Collaboration],
  Phys.\ Rev.\ Lett.\  {\bf 87}, 071301 (2001).
\bibitem{SNO2}
  Q.~R.~Ahmad {\it et al.} [SNO Collaboration],
  Phys.\ Rev.\ Lett.\  {\bf 89}, 011301 (2002)
  and {\it ibid.} {\bf 89}, 011302 (2002).
\bibitem{SNO3}
  S.~N.~Ahmed {\it et al.} [SNO Collaboration],
  Phys.\ Rev.\ Lett.\  {\bf 92}, 181301 (2004).
\bibitem{SNOsalt}
  B.~Aharmim {\it et al.}  [SNO Collaboration],
  Phys.\ Rev.\ C {\bf 72}, 055502 (2005).
\bibitem{SKatm}
  Y.~Ashie {\it et al.}  [Super-Kamiokande Collaboration],
  Phys.\ Rev.\ D {\bf 71}, 112005 (2005).
\bibitem{KamLAND}
  K.~Eguchi {\it et al.}  [KamLAND Collaboration],
  Phys.\ Rev.\ Lett.\  {\bf 90}, 021802 (2003).

\bibitem{K2K}
  M.~H.~Ahn  [K2K Collaboration],
  hep-ex/0606032.
\bibitem{LSND}
  A.~Aguilar {\it et al.} [LSND Collaboration],
  Phys.\ Rev.\ D {\bf 64}, 112007 (2001).


\bibitem{miniboone}
  A.~A.~Aguilar-Arevalo {\it et al.} [MiniBooNE Collaboration],\\ 
  The MiniBooNE Run Plan, available at

  \url{http://www-boone.fnal.gov/publicpages/runplan.ps.gz}


\bibitem{BPont57}
  B.~Pontecorvo,
  Sov.\ Phys.\ JETP {\bf 6}, 429 (1957)
  [Zh.\ Eksp.\ Teor.\ Fiz.\  {\bf 33}, 549 (1957)] 
  and {\it ibid.} {\bf 7}, 172 (1958) [{\it ibid.} {\bf 34} (1958)
  247];
  Z.~Maki, M.~Nakagawa and S.~Sakata,
  Prog.\ Theor.\ Phys.\ {\bf 28}, 870 (1962).


\bibitem{CHOOZ}
  M.~Apollonio {\it et al.}  [CHOOZ Collaboration],
  Phys.\ Lett.\ B {\bf 466}, 415 (1999).

\bibitem{PaloV}
  F.~Boehm {\it et al.},
  Phys.\ Rev.\ Lett.\ {\bf 84}, 3764 (2000);
  and Phys.\ Rev.\ D {\bf 62}, 072002 (2000).
\bibitem{MINOS}
  E.~Ables {\it et al.} [MINOS Collaboration],
  FERMILAB-PROPOSAL-0875.
\bibitem{MINOSRECENT}
D.~G.~Michael {\it et al.}  [MINOS Collaboration],
  Phys.\ Rev.\ Lett.\  {\bf 97}, 191801 (2006).

\bibitem{thomas}
T.~Schwetz,
  Phys.\ Scripta {\bf T127}, 1 (2006).
\bibitem{newfit}
G.~L.~Fogli {\it et al.},
  arXiv:hep-ph/0608060.
\bibitem{KL766}
  T.~Araki {\it et al.}  [KamLAND Collaboration],
  Phys.\ Rev.\ Lett.\  {\bf 94}, 081801 (2005).
\bibitem{geer}
S.~Geer,
 Phys.\ Rev.\ D {\bf 57}, 6989 (1998)
[Erratum-ibid.\ D {\bf 59}, 039903 (1999)].
\bibitem{nf1}
A.~De Rujula, M.~B.~Gavela and P.~Hernandez,
  Nucl.\ Phys.\ B {\bf 547}, 21 (1999).

\bibitem{nf2}
V.~D.~Barger, S.~Geer and K.~Whisnant,
\bibitem{nf3}
A.~Donini {\it et al}, 
  Nucl.\ Phys.\ B {\bf 574}, 23 (2000).
\bibitem{nf4}
V.~D.~Barger {\it et al}, 
  Phys.\ Rev.\ D {\bf 62}, 013004 (2000).
\bibitem{nf5}
V.~D.~Barger {\it et al}, 
  Phys.\ Rev.\ D {\bf 62}, 073002 (2000).

\bibitem{nf6}
A.~Cervera, {\it et al}
  Nucl.\ Phys.\ B {\bf 579}, 17 (2000)
  [Erratum-ibid.\ B {\bf 593}, 731 (2001)].

\bibitem{nf6b}

M.~Freund, P.~Huber and M.~Lindner,
  Nucl.\ Phys.\ B {\bf 585}, 105 (2000).

\bibitem{nf7}
 V.~D.~Barger {\it et al}, 
  Phys.\ Lett.\ B {\bf 485}, 379 (2000).
\bibitem{nf7b}
J.~Burguet-Castell {\it et al}, 
  Nucl.\ Phys.\ B {\bf 608}, 301 (2001).
\bibitem{nf7c}
M.~Freund, P.~Huber and M.~Lindner,
  Nucl.\ Phys.\ B {\bf 615}, 331 (2001).

\bibitem{silver}
  A.~Donini, D.~Meloni and P.~Migliozzi,
  Nucl.\ Phys.\ B {\bf 646}, 321 (2002);

  D.~Autiero {\it et al.},
  Eur.\ Phys.\ J.\ C {\bf 33}, 243 (2004).

\bibitem{study1-physics}
C.~Albright {\it et al.},
  arXiv:hep-ex/0008064;


\bibitem{nf8}
A.~Blondel {\it et al.},
  Nucl.\ Instrum.\ Meth.\ A {\bf 451}, 102 (2000);

M.~Apollonio {\it et al.},
  arXiv:hep-ph/0210192;

C.~Albright {\it et al.}  [Neutrino Factory/Muon Collider Collaboration],
arXiv:physics/0411123.
\bibitem{yo}
O.~Mena,
  Mod.\ Phys.\ Lett.\ A {\bf 20}, 1 (2005).
\bibitem{nf9}
  P.~Huber, M.~Lindner, M.~Rolinec and W.~Winter,
  hep-ph/0606119.

\bibitem{zucc}
  P.~Zucchelli,
  Phys.\ Lett.\ B {\bf 532}, 166 (2002).


\bibitem{mauro}
  M.~Mezzetto,
  J.\ Phys.\ G {\bf 29}, 1781 (2003)
  and {\it ibid.} {\bf 29} 1771 (2003).

\bibitem{betabeam1}
  J.~Burguet-Castell {\it et al.},
  Nucl.\ Phys.\ B {\bf 695}, 217 (2004)

\bibitem{betabeam2}
A.~Donini {\it et al.},
  Nucl.\ Phys.\ B {\bf 710}, 402 (2005).
\bibitem{betabeam3}
  J.~Burguet-Castell {\it et al.},
  Nucl.\ Phys.\ B {\bf 725}, 306 (2005).

\bibitem{betabeam4}
  P.~Huber {\it et al.},
  hep-ph/0506237.
\bibitem{betabeam5}
A.~Donini and E.~Fernandez-Martinez,
arXiv:hep-ph/0603261.
\bibitem{tabarellis1}
 A.~Donini {\it et al.},
  hep-ph/0604229.
\bibitem{study1}
N. Holtkamp and D. Finley, eds.,
FERMILAB-PUB-00/108-E
\bibitem{study2}
S. Ozaki {\it et al.},
BNL-52623,2001
\bibitem{study2a}
S. Geer and M. Zisman, eds.,
BNL-72369-2004, FERMILAB-TM-2259, LBNL-55478;\\
S. Geer and M.S. Zisman,
FERMILAB-PUB-06-454-E, Submitted to Prog. Part. Nucl. Phys.
\bibitem{Nova}
D.~Ayres {\it et al.},
Oscillations in the Fermilab NuMI Beamline,''
hep-ex/0503053.
\bibitem{nufact06-det}   
M. Ellis, "ISS Detector Working Group Report", presented at the 8th International Workshop on Neutrino Factories, Superbeams and Betabeams (NUFACT06), UC Irvine, August 24-30,2006, http://nufact06.physics.uci.edu/Default.aspx
\bibitem{FL96} 
G.~L.~Fogli and E.~Lisi,
  Phys.\ Rev.\ D {\bf 54}, 3667 (1996).
\bibitem{MN01}
  H.~Minakata and H.~Nunokawa,
  JHEP {\bf 0110}, 001 (2001).
\bibitem{BMWdeg} 
V.~D.~Barger, S.~Geer, R.~Raja and K.~Whisnant,
  Phys.\ Rev.\ D {\bf 63}, 113011 (2001).
\bibitem{deg}
  T.~Kajita, H.~Minakata and H.~Nunokawa,
  Phys.\ Lett.\ B {\bf 528}, 245 (2002);
  H.~Minakata, H.~Nunokawa and S.~J.~Parke,
  Phys.\ Rev.\ D {\bf 66}, 093012 (2002);
  P.~Huber, M.~Lindner and W.~Winter,
  Nucl.\ Phys.\ B {\bf 645}, 3 (2002);
  A.~Donini, D.~Meloni and S.~Rigolin,
  JHEP {\bf 0406}, 011 (2004);
  M.~Aoki, K.~Hagiwara and N.~Okamura,
  Phys.\ Lett.\ B {\bf 606}, 371 (2005);
  O.~Yasuda,
  New J.\ Phys.\  {\bf 6}, 83 (2004);
  O.~Mena and S.~J.~Parke,
  Phys.\ Rev.\ D {\bf 72}, 053003 (2005).
\bibitem{MN97}
  H.~Minakata and H.~Nunokawa,
  Phys.\ Lett.\ B {\bf 413}, 369 (1997).
\bibitem{BMW02off}
  V.~Barger, D.~Marfatia and K.~Whisnant,
  Phys.\ Rev.\ D {\bf 66}, 053007 (2002).
\bibitem{SN1}
  O.~Mena Requejo, S.~Palomares-Ruiz and S.~Pascoli,
  Phys.\ Rev.\ D {\bf 72}, 053002 (2005).
\bibitem{twodetect}
  M.~Ishitsuka {\it et al.}, 
  Phys.\ Rev.\ D {\bf 72}, 033003 (2005);
  K.~Hagiwara, N.~Okamura and K.~i.~Senda,
  Phys.\ Lett.\ B {\bf 637}, 266 (2006).
\bibitem{SN2}
O.~Mena, S.~Palomares-Ruiz and S.~Pascoli,
  Phys.\ Rev.\ D {\bf 73}, 073007 (2006).
\bibitem{T2kk}
  T.~Kajita {\it et al},
  arXiv:hep-ph/0609286.
\bibitem{otherexp1}
  J.~Burguet-Castell {\it et al.},
  Nucl.\ Phys.\ B {\bf 646}, 301 (2002);
\bibitem{HLW02}
  P.~Huber, M.~Lindner and W.~Winter,
  Nucl.\ Phys.\ B {\bf 654}, 3 (2003).
\bibitem{MNP03}
  H.~Minakata, H.~Nunokawa and S.~J.~Parke,
  Phys.\ Rev.\ D {\bf 68}, 013010 (2003).
\bibitem{BMW02}
  V.~Barger, D.~Marfatia and K.~Whisnant,
  Phys.\ Lett.\ B {\bf 560}, 75 (2003).
\bibitem{otherexp}
  K.~Whisnant, J.~M.~Yang and B.~L.~Young,
  Phys.\ Rev.\ D {\bf 67}, 013004 (2003);
  P.~Huber {\it et al.},
  Nucl.\ Phys.\ B {\bf 665}, 487 (2003);
  P.~Huber {\it et al.},
  Phys.\ Rev.\ D {\bf 70}, 073014 (2004);
  A.~Donini, E.~Fern\'andez-Mart\'{\i}nez and S.~Rigolin,
  Phys.\ Lett.\ B {\bf 621}, 276 (2005).
\bibitem{mp2}
  O.~Mena and S.~J.~Parke,
  Phys.\ Rev.\ D {\bf 70}, 093011 (2004).
\bibitem{HMS05}
  P.~Huber, M.~Maltoni and T.~Schwetz,
  Phys.\ Rev.\ D {\bf 71}, 053006 (2005).
\bibitem{huber2}
A.~Blondel {\it et al.},
  hep-ph/0606111.
\bibitem{lastmine}
O.~Mena, H.~Nunokawa and S.~J.~Parke,
  hep-ph/0609011;
O.~Mena,
  hep-ph/0609031.
\bibitem{dis}
  A.~Bueno, M.~Campanelli and A.~Rubbia,
  Nucl.\ Phys.\ B {\bf 589} 577 (2000). 
\bibitem{Stef}
  A.~Donini {\it et al}
  Nucl.\ Phys.\ B {\bf 743}, 41 (2006).
\bibitem{Akhmedov:2004ny}
E.~K.~Akhmedov {\it et al.},
JHEP {\bf 0404} 078 (2004).



\end{thebibliography}
\end{document}